\documentclass[review,12pt,authoryear,times]{elsarticle}

\usepackage{array}

\usepackage{natbib}
\usepackage{color}
\usepackage[latin1]{inputenc}
\usepackage{graphicx}
\usepackage{rotating}
\usepackage{booktabs}
\usepackage{amsmath}
\usepackage{epstopdf}
\usepackage{amssymb}
\usepackage{overpic}
\usepackage{booktabs}
\journal{Icarus}
\begin{document}
\begin{frontmatter}
\title{Comets formed in solar-nebula instabilities! -- An experimental and
modeling attempt to relate the activity of comets to their formation process}
\author[igep]{J. Blum}
\author[igep]{B. Gundlach}
\author[igep]{S. Mühle}
\author[ieec]{J.M. Trigo-Rodriguez}
\address[igep]{Institut für Geophysik und extraterrestrische Physik, Technische Universität Braunschweig, \\Mendelssohnstr. 3, D-38106 Braunschweig, Germany}
\address[ieec]{Institute of Space Sciences (CSIC), Campus UAB, Facultat de Ciències, \\Torre C-5 pares, 2a pl., 08193 Bellaterra (Barcelona), Spain}
\begin{abstract}
When comet nuclei approach the Sun, the increasing energy flux through the surface layers leads to sublimation of the underlying ices and subsequent outgassing that promotes the observed emission of gas and dust. While the release of gas can be straightforwardly understood by solving the heat-transport equation and taking into account the finite permeability of the ice-free dust layer close to the surface of the comet nucleus, the ejection of dust additionally requires that the forces binding the dust particles to the comet nucleus must be overcome by the forces caused by the sublimation process. This relates to the question of how large the tensile strength of the overlying dust layer is. Homogeneous layers of micrometer-sized dust particles reach tensile strengths of typically $10^3$ to $10^4$ Pa. This exceeds by far the maximum sublimation pressure of water ice in comets. It is therefore unclear how cometary dust activity is driven.
\par
To solve this paradox, we used the model by Skorov and Blum (Icarus 221, 1-11, 2012), who assumed that cometesimals formed by gravitational instability of a cloud of dust and ice aggregates and calculated for the corresponding structure of comet nuclei tensile strength of the dust-aggregate layers on the order of 1 Pa. Here we present evidence that the emitted cometary dust particles are indeed aggregates with the right properties to fit the model by Skorov and Blum. Then we experimentally measure the tensile strengths of layers of laboratory dust aggregates and confirm the values derived by the model. To explain the comet activity driven by the evaporation of water ice, we derive a minimum size for the dust aggregates of $\sim 1$ mm, in agreement with meteoroid observations and dust-agglomeration models in the solar nebula. Finally we conclude that cometesimals must have formed by gravitational instability, because all alternative formation models lead to higher tensile strengths of the surface layers.

\end{abstract}
\end{frontmatter}

\setlength{\tabcolsep}{10pt}
\renewcommand{\arraystretch}{2}
\renewcommand{\topfraction}{1.0}
\renewcommand{\bottomfraction}{1.0}
\section{\label{sec:introduction}Introduction: formation scenarios of planetesimals and cometesimals}
It is now well established that dust inside the snow line of the solar nebula quickly coagulated into millimeter- to centimeter-sized agglomerates due to direct sticking in collisions \citep{Guettler2010,Zsom2010}. The further growth to planetesimal-sized objects is still under debate, with two major scenarios under consideration: the mass transfer scenario (1) and the gravitational instability scenario (2).
\par
(1) As direct sticking is mostly prevented by bouncing among the dust aggregates \citep{BlumMuench1993,Langkowski2008,Weidling2009,Beitz2012,Weidling2011,Schraepler2012,Deckers2013}, only those particles colliding with velocities slower than the sticking-bouncing transition can further grow, whereas the fastest collisions in the ensemble lead to fragmentation with mass transfer \citep{Windmark2012,Windmark2012b,Garaud2013}. This latter process has been extensively studied in the laboratory \citep{Wurm2005,Teiseretal2009a,Teiseretal2009b,Guettler2010,Kothe2010,Teiseretal2011} and is now well established for aggregates consisting of micrometer-sized silicate grains. It has been shown that in principle planetesimals can form by this process \citep{Windmark2012,Windmark2012b,Garaud2013} although the timescales are rather long and details about counteracting processes \citep[e.g., erosion;][]{Schraepler2011} need to be clarified.
\par
(2) Alternatively, \citet{Johansen2007} showed that cm-sized particles can be sufficiently concentrated by the streaming instability \citep{Youdin2005} so that the ensemble becomes gravitationally unstable and forms planetesimals. Also here, several details need to be clarified before this process can be regarded as established, e.g. the collisional fate of the dust agglomerates within the instabilities, fragmentation of the collapsing cloud and the mass distribution function of the resulting planetesimals, and the required high metallicity of the solar nebula.
\par
In the outer solar nebula beyond the snow line, the dominant material should be (water) ice. Due to the higher anticipated stickiness of water-ice particles \citep{Gundlach2011b}, ice aggregates are supposed to grow to larger masses and fluffier structures in the outer solar nebula \citep{Wada2008,Wada2009,Okuzmi2012,Kataokaetal2013}. As empirical proof for this concept from laboratory experiments is still missing, it can only be speculated whether icy planetesimals form directly by hit-and-stick collisions, or by a multi-step process. If direct formation of cometesimals is not feasible, similar processes as discussed above for the inner solar nebula might also apply for its outer reaches.
\par
Since the first space missions to comet Halley it has been known that comets consist in almost equal parts of ice and refractory materials (dust), with the addition of organic materials \citep{Jessbergeretal1988}, which in turn led to revised cometary dust modeling \citep{Greenberg1990,Li1997,Greenberg1998}. The samples brought back from comet 81P/Wild by the Stardust mission revealed that the refractory materials are high-temperature condensates, which must have been radially mixed outwards before the formation of the comet nucleus \citep{McKeegan2006,Zolensky2006}. As the growth timescales to mm or cm sizes are rather short in the inner solar nebula (a few $10^3$ years) and as the dust aggregates are supposed to be rather compact (with a porosity of ``only'' $60-65 \, \%$, according to \citet{Zsom2010} and \citet{Weidling2009}), with any further growth slowed down due to the decreased stickiness of large dust aggregates \citep{Guettler2010}, it is plausible to assume that the refractory materials were mixed into the outer solar nebula in form of mm- to cm-sized agglomerates (see also Sects. \ref{sec:comcomp} and \ref{sec:comtensile}). Hence, cometesimals in the outer solar nebula were then formed out of icy and dusty agglomerates by one of the two processes described above, namely (1) fragmentation with mass transfer (MT) or (2) spatial enhancements in (magneto-)hydrodynamic instabilities with subsequent gravitational instability (GI). From this line of reasoning, we can derive several physical distinctions in the resulting icy-dusty planetesimals. We summarize these in the Table \ref{tab:comprop}.
\begin{table*}
  \centering
  \caption{Comparison between the two formation scenarios of icy-dusty
  planetesimals. GI stands for gravitational instability, MT represents the
  process of mass transfer.}\label{tab:comprop}
        \scriptsize   
        \vspace{2mm}
        \begin{tabular}{lcccc}
        \hline
        & GI & & MT & \\
        \hline
        Volume filling factor & $0.35 \times 0.6 \approx 0.2$ &[1,7] & $\sim$0.4 &[2]  \\
        Tensile strength of interior [Pa] & $\sim 10$ &[3] & $\sim 10000$ &[2,5] \\
        Tensile strength of ice-free outer dust layer [Pa] & $\sim 1$ &[3] & $\sim 1000$ &[2,5]   \\
        Gas permeability [$\rm m^4 s^{-1}$] & $\sim 1 \times 10^{-6}$ &[4] & $\sim 1 \times 10^{-9}$& [4]  \\
        Thermal conductivity [$\rm W m^{-1} K^{-1}$]  & $10^{-3}-1$ &[6] & $10^{-2}-10^{-1}$& [6]  \\
        & (conduction/radiation)& & (conduction) &\\
        \hline
\multicolumn{4}{l}{References:}\\[-0.2cm]
\multicolumn{4}{l}{[1] \citet{Weidling2009}, [2] \citet{Kothe2010}, [3] \citet{Skorov2012}, [4] \citet{Gundlach2011}, [5] \citet{Blum2006a},}\\[-0.2cm]
\multicolumn{4}{l}{[6] \citet{Gundlach2012}, [7] \citet{Zsom2010}.}
\end{tabular}
\end{table*}
It should be mentioned that we assume that the formation process for cometesimals was the same anywhere in the outer solar nebula. Thus, the following discussion in this paper refers to both, Kuiper-belt and Oort-cloud comets. As to the formation timescales for cometesimals, these are required to be long enough for the radial mixing of the high-temperature condensates to occur, but certainly shorter that the lifetime of the nebula gas. As this might be a problem for the MT origin of cometesimals at large heliocentric distances, the timescales for the instability-driven formation of cometesimals should always be sufficiently short. In the latter, however, the aggregate sizes at which the bouncing barrier is reached and for which then some concentration process forms a gravitationally unstable cloud, could be considerably different (albeit yet unknown) for the two reservoirs of Kuiper-belt and Oort-cloud comets.
\par
The volume filling factor $\phi$ is defined as the fraction of the total volume occupied by the material and is related to the porosity $\psi$ by $\phi \, = \, 1 \,  -\, \psi$. For an icy-dusty planetesimal formed by the GI process, the volume filling factor is determined by the packing fraction of the dust aggregates into the planetesimal ($\phi_{\rm global} \approx 0.6$, if we assume that the dust aggregates pack almost as densely as possible) and by the volume filling factor of the individual dust aggregates \citep[$\phi_{\rm local} \approx 0.35$, according to][]{Weidling2009}. The volume filling factor of the MT dust aggregates has been measured to be close to $\phi_{\rm local} = 0.4$ \citep{Kothe2010}. The tensile strength of a package of dust aggregates, which collapsed under their own gravity to form a km-sized body with a volume filling factor of $\phi_{\rm global}$ has been calculated by \citet{Skorov2012} to be
    \begin{equation}\label{Eq:tsmodel}
        p_{\rm tensile} \, = \, p'_{\rm tensile} \phi_{\rm global} \left( \frac{s}{\mathrm{1 mm}} \right)^{-2/3} ,
    \end{equation}
with $p'_{\rm tensile}=1.6 \, \mathrm{Pa}$ and $s$ being the radius of the infalling dust aggregates. For ice aggregates, the tensile strength is supposed to be a factor of ten higher \citep{Gundlach2011b}. In the case of planetesimals formed by the MT process, their rather compact packing of the monomer grains ensures a relatively higher tensile strength of $\sim 1 \, \mathrm{kPa}$ for volatile-free and $\sim 10 \, \mathrm{kPa}$ for icy particles \citep{Blum2006a}. Due to the smaller pore size in the planetesimals formed by MT (the pore size is on the order of the monomer-grain size, i.e., $\sim 1 \, \mathrm{\mu m}$, whereas for planetesimals formed by GI the pore size is on the order of the aggregate size), the gas permeability is much lower \citep{Gundlach2011}. The thermal conductivity is not easily distinguishable between the two formation models, due to the fact that for large pore sizes, radiative energy transport is no longer negligible \citep{Gundlach2012}. Thus, the range of possible thermal-conductivity values is much larger for the GI-formed planetesimals than for those formed by MT.
\par
As mentioned above, \citet{Skorov2012} were the first to bring up the distinction in tensile strength between the two models, who related the formation of icy-dusty planetesimals to present comet nuclei, and who showed that, according to their model (see Eq. \ref{Eq:tsmodel}), only the GI model can explain a continued gas and dust activity of a comet. Their model for the tensile strength of the ice-free outer layers of a comet nucleus is based on the assumption that dust and ice aggregates once formed the comet nucleus by gravitational instability so that essentially the aggregates collapsed below or at the very low escape speed of the kilometer-sized body. Thus, the aggregates are only slightly deformed and the resulting binding between the clumps is much weaker than in the mass-transfer process.
\par
In this article, we intend to verify the model by \citet{Skorov2012} and to support their statement that comets were formed in gravitational instabilities. This will be done in the following: in Sect. \ref{sec:comcomp} we will show that comet nuclei indeed consist of mm- to cm-sized dust particles and ice clumps with at least these sizes. In Sect. \ref{sec:comtensile}, we will then show that observed cometary dust aggregates or meteoroids are consistent with model expectations for dust aggregates in the bouncing regime, i.e. with a rather low porosity and a correspondingly rather low tensile strength $p_{\rm tensile}$ (in this paper, we denote $p_{\rm tensile} \sim 1~\mathrm{kPa}$ as low tensile strength and $p_{\rm tensile} \sim 1~\mathrm{Pa}$ as ultra-low tensile strength)\footnote{Here, it should be mentioned that the internal tensile strength of aggregates and meteoroids is low ($p_{\rm tensile} \sim 1-10~\mathrm{kPa}$), whereas an arrangement of aggregates possess an ultra-low tensile strength ($p_{\rm tensile} \sim 1~\mathrm{Pa}$).}. In the following experimental part (see Sect. \ref{sec:lab}), we will then construct an analog sample of a cometary ice-free surface and measure its tensile strength. The results of our laboratory experiments and their analysis will then be applied to comet nuclei (see Sect. \ref{sec:application}). We conclude in Sect. \ref{sec:conclusion} that comets were formed by gravitational instability and will give a short outlook of future work in Sect. \ref{sec:future}.

\section{\label{sec:comcomp}Comets consist of large and porous dust and ice aggregates}
The basis of the comet-nucleus formation model by \citet{Skorov2012} is the existence of at least mm-sized dust and ice agglomerates, which collapse due to gravitational instability at relatively low velocities and, hence, form a weakly bound (both, gravitationally as well as cohesively) cometesimal. In the following, we will provide evidence that these dusty and icy clumps indeed exist close to the surfaces of comet nuclei.
\par
There is observational evidence that mm to dm-sized particles exist in the coma and the trail of comet 2P/Encke \citep{Reachetal2000}. Additionally, Spitzer observations of the dust trail of comet 67P/Churyumov-Gerasimenko have shown that the dust trail consists of large particles bigger than $100 \, \mathrm{\mu m}$ \citep{Agarwaletal2010}. Neckline observations of comet 67P/Churyumov-Gerasimenko by \citet{Ishiguro2008} show the presence of cm-sized particles. There is also evidence that the release of cm-sized particles from active regions and their subsequent sublimation and fragmentation induces the outburst of Centaur-like comet 29P/Schwassmann-Wachmann 1 \citep{Trigoetal2010}. Large dust particles were also found for comet C/1995 O1 \citep[Hale-Bopp;][]{Gruenetal2001} and Arecibo observations of comet C/2001 A2 (LINEAR) show the existence of $\sim 2 \, \mathrm{cm}$ grains \citep{Nolanetal2006,Harmonetal2004}. The cameras of the EPOXI mission saw the ejection of dm-sized ice particles from the surface of comet 103P/Hartley \citep{AHearnetal2011}. This was the first time that direct evidence of granular ice was found in a comet nucleus. A detailed analysis of the particle sizes yielded a lower size limit of $\gtrsim 1 \, \mathrm{cm}$ and a very steep size distribution with power-law slopes ranging from $-6.6$ to $-4.7$ \citep{Kelleyetal2013}. Due to the spatial distribution of the particles, the authors concluded that the particles are porous ice aggregates with densities ranging from $1 \, \mathrm{kg \, m^{-3}}$ to $1000 \, \mathrm{kg \, m^{-3}}$. Furthermore, the on board dust detector of the Stardust spacecraft measured particle sizes up to $2 \, \mathrm{mm}$ in the coma of comet 81P/Wild \citep{Tuzzolinoetal2004}. Additionally, cometary dust particles were collected during the Stardust mission by an aerogel dust collector. The dimensions of the impact craters in the aerogel were analyzed by \citet{Burchelletal2008} in order to get an idea of the size of the cometary dust particles before the impact. This investigation has shown that the dimensions of the craters in the aerogel are ranging from $\mathrm{\mu m}$- to $\mathrm{mm}$, which indicates that large dust particles were present in the coma of comet 81P/Wild during the Stardust mission.
\par
It was suggested by \citet{Kolokolovaetal2004} that the macroscopic particles in cometary comae are aggregates of smaller grains because the particle size shows a dependency on radial distance to the nucleus. \citet{Sykesetal2004} review meteoroid densities and find for long-period comets values of $200-600 \, \mathrm{kg\, m^{-3}}$; these extremely low densities (which imply volume filling factors of $\phi_{\rm global} \approx 0.1$ to $0.3$) also suggest that meteoroids are conglomerates of dust aggregates.
\par
Thus, it is evident that comet nuclei consist in a significant proportion of macroscopic agglomerates of dust and ice. However, a fair fraction of the comet material could also consist of microscopic particles. The ratio of macroscopic agglomerates to microscopic particles is unknown, but future in-situ measurements like the Rosetta mission, will help to constrain this.
\par
In the following section, we will show that the inner tensile (i.e., cohesive) strength of these aggregates is well above the outgassing pressure of a comet nucleus so that the observed aggregates cannot be simply fragments ripped off a larger dusty body by gas pressure, but are required to be loosely held together.

\section{\label{sec:comtensile}The tensile strength of meteor-shower particles}
In our modern view, cometary materials were assembled in the outer regions of the protoplanetary disk (see above), at heliocentric distances large enough to allow ices and organic materials to condense. Comets release dust particles present in their interiors thanks to the sublimation of ices heated by solar irradiation. This mechanism was first envisioned and described by Fred L. Whipple as the cause of the formation of dense meteoroid streams, following similar orbits to their progenitor comets \citep{Whipple1950,Whipple1951}. The outflowing gas drags grains away from the comet, but usually the coupling between gas and dust is poor. Even if the ice sublimation is important and the flux of ejected gas is relevant, only a fraction of the energy is transferred to the dust. In general, the smaller the particle, the more efficient is the coupling between dust and gas \citep{Jenniskens2006}. Consequently, the comet dust tail is an ensemble of small solid particles recently dragged away by the gas outflow and affected by the solar radiation pressure, but also subjected to solar gravity. In contrast, the meteoroid stream released from a comet consists of dust particles so large that their relative velocities to the comet nucleus are small and radiation pressure is relatively unimportant on short timescales. Thus, the stream particles will be forced to follow a heliocentric orbit according to Kepler's laws with orbital elements similar to those of the comet.
\par
Most cometary particles reaching the Earth are associated with these streams. In general, the dust particles ablate when they penetrate into the atmosphere at high geocentric velocities. Ablation is a mass-loss process in which the meteoroid particles decelerate due to the collisions and friction with atmospheric components. It is known that a significant fraction of meteoroids (increasing with increasing size) are fragile dust aggregates that easily break apart when the experienced stress exceeds the strength limit of the material \citep{TrigoLlorca2006,TrigoLlorca2007}. Basically, the tensile strength of these particles is the maximum amount of tensile stress that they can be subjected to before they break. The aerodynamic strength of cometary meteoroids can be derived using the equation
\begin{equation}\label{Eq:aerstr}
    p_{\rm aero} \, = \,  \rho_{\textrm atm} \, v^2
\end{equation}
\citep{Bronshten1981}, where $\rho_{\mathrm atm}$ and $v$ are the atmospheric mass density at the height where the meteoroid breaks up and the velocity of the particle at this point, respectively. If the density is given in units of $\rm kg \,m^{-3}$ and the velocity in $\rm m \, s^{-1}$, the strength is given in Pa.
\par
\citet{Jacchia1958} and \citet{Ceplecha1958} identified three populations of meteoroids that exhibit low tensile strengths and have orbital elements that are clearly associated with comets. \citet{Verniani1969}, \citet{Verniani1973} and \citet{Millman1972} found that most of the sporadic meteoroids of cometary origin are highly porous and fragment when the pressure exceeds $2 \times 10^3 \, \mathrm{Pa}$. Such disruptive-strength values are characteristic for most of the stream meteors studied by \citet[][see also Table \ref{tab:metstreams}]{TrigoLlorca2006} and are consistent with recent published data for other minor streams. We searched in the scientific literature the available data on the dynamic tensile strength of meteoroids and Table \ref{tab:metstreams} compiles the available data. We also searched the literature in order to obtain accurate trajectory and velocity data of meteor showers and applied Eq. \ref{Eq:aerstr} to compute the aerodynamic strengths of the meteoroids from the average values of the height of shower meteors, considering the observed velocity and the atmospheric density at the brightest point of the trajectory. The latter was derived from the U.S. standard atmosphere\footnote{U.S. standard atmosphere (1976 version).}. We used the observed velocities rather than the geocentric ones, because we are dealing with the effect of stress produced by the collision of the meteoroids with atmospheric components. Table \ref{tab:metstreams} shows the available data.
\begin{table*}
  \centering
  \caption{Known meteor streams, their cometary sources, the inferred tensile strengths of the dust particles following Eq. \ref{Eq:aerstr}, the number of meteors in the sample, the references of the individual observations, and the volume filling factor of the meteoroids measured by \citet{BabadzhanovKokhirova2009}.}\label{tab:metstreams}
       \scriptsize 
        \vspace{2mm}
  \begin{tabular}{llcccc}
    \hline
    Meteoroid & Parent & Disruption Strength & Number of & Volume Filling & Reference\\
    Stream    & Body   & ($\times 10^3 \mathrm{Pa}$) & Meteors & Factor\\
    \hline
    Taurids	& 2P/Encke & $34 \pm 7$ & 14  & $0.59 \pm 0.19$ & [1]\\
    Geminids & (3200) Phaeton & $22 \pm 2$ & 196  & $1.00 \pm 0.25$ & [1]\\
    Quadrantids	& 2003 EH1 & $\sim 20$ & 39  & $0.56 \pm 0.19$ & [1]\\
    Perseids & 109P/Swift-Tuttle & $12 \pm 3$ & 112  & $0.55 \pm 0.01$ & [1]\\
    Orionids & 1P/Halley & $6 \pm 3$ & 3  & $0.38 \pm 0.24$ & [1]\\
    Aquarids & unknown & $12 \pm 3$ & 5  & $0.71 \pm 0.26$ & [1]\\
    Leonids  & 55P/Tempel-Tuttle & $6 \pm 3$ & 24  & $0.17 \pm 0.06$ & [1]\\
    October Draconids & 21P/Giacobini-Zinner & $0.4 \pm 0.1$ & 52  & 0.17* &[1,2]\\
    Hydrids & unknown & $19 \pm 4$ & 1  & - & [3]\\
    Omicron Cygnids & unknown & $19 \pm 3$ & 1  & $0.88 \pm 0.72$ &[4]\\
    JFC meteoroid & unknown JFC & $21 \pm 7$ & 1  & - & [5]\\
    Comae Berenicids & unknown & $150 \pm 70$ & 1  & - & [6]\\
    Capricornids & 141P/Machholz, & $23 \pm 3$ & 9  & $0.62 \pm 0.15$ & [1,2,8]\\
    & 45P/Honda-Mrkos-&&&&\\
    & Pajdusakova, &&&&\\
    & or (9162) 1987 OA	&&&&\\
    \hline
    \multicolumn{6}{l}{References:}\\[-0.5cm]
    \multicolumn{6}{l}{[1] \citet{TrigoLlorca2006}, [2] \citet{Trigo2013}, [3] \citet{Robles2013}, [4] \citet{Jimenez2013},}\\[-0.5cm]
    \multicolumn{6}{l}{[5] \citet{Diez2012}, [6] \citet{Martinez2012},[7] \citet{Rodriguez2012}, [8] \citet{Zamorano2012}.}\\[-0.5cm]
    \multicolumn{6}{l}{*: Data taken from \citet{Borovicka2007} (no information about the uncertainty of the porosity is available).}
  \end{tabular}
\end{table*}
\par
From the meteor data, the aerodynamic tensile strength can be estimated with relatively good accuracy, following Eq. \ref{Eq:aerstr}. The only two properties entering the equation, the meteor velocity and atmospheric density, can be derived with sufficient precision. Typically, the velocities at the disruption height can be calculated from multiple-station meteor observations from the ground with an accuracy of $0.2 \, \mathrm{km \, s^{-1}}$ to $0.6\, \mathrm{km \, s^{-1}}$. To compute the aerodynamic strength by using Eq. \ref{Eq:aerstr} requires the exact knowledge of the height where the meteoroid fragments. For cometary meteoroids, this instant is associated with a sudden increase in the meteor luminosity that is produced immediately after the break-up \citep{TrigoLlorca2006,TrigoBlum2009}. As a consequence of the quick release of dust grains, the effective cross section of the meteoroid and also the mass available for ablation increase, both increasing the meteor luminosity. The break-up produces a distinctive flare at a well-defined and measurable height. Accurate multiple-station observing programs are required to measure meteor heights and velocities. However, not all meteors suffer a clear disruption, but those that do provide us with direct information about the tensile strength of the comet material. The results presented in Table \ref{tab:metstreams} clearly show that most of the cometary meteoroids that are reaching Earth's atmosphere exhibit tensile strengths of $\sim 10^4 \, \mathrm{Pa}$ with deviations of up to a factor of ten for individual comets.
\par
The data shown in Table \ref{tab:metstreams} stem from meteoroids typically up to 1 cm in diameter, as suggested by the dynamic masses derived from the study of fireballs produced by cometary meteoroids \citep[see, e.g.,][]{Rietmeijer2000}. It is obvious that these sizes set constraints for $s$ in Eq. \ref{Eq:tsmodel}, which are in full agreement with the model by \citet{Zsom2010}, which predicts typical dust-aggregate sizes in the inner parts of the young Solar System of millimeters to centimeters. Mind, however, that the detection of meteoroids in the visual range, as discussed in Table \ref{tab:metstreams}, depends mostly on their ability to ionize light, which is a function of their incoming kinetic energy \citep{Jenniskens2006}. Thus, less massive bodies will most likely be overlooked. On top of that, the lifetimes of these bodies on the meteoroid-stream orbits is limited by Poynting-Robertson drag so that the meteor method is biased towards larger bodies.
\par
The study of the ablation of meteoroids in the atmosphere can provide additional clues on their porosities. \citet{GustafsonAdolfsson1996} demonstrated that in favorable cases the bulk density and porosity of meteoroids during atmospheric flight can be estimated. By using the data obtained during ablation of the European Network type IIIb bolide (EN 71177), they estimated a bulk density of $260 \, \mathrm{kg \, m^{-3}}$. From this result, they obtained the packing factor by calculating the ratio of the density of the compact material of chondritic composition (assumed to have $2400 \, \mathrm{kg \, m^{-3}}$) to the bulk density. They estimated that the meteoroid was $\sim700\, \mathrm{g}$ in mass, having $\sim 88 \,\%$ porosity, i.e., a volume filling factor of $\phi_{\rm global} \, = \, 0.12$. This low packing fraction is in full accord with the minimum values of volume filling factors estimated for IDPs \citep{Rietmeijer1998,Rietmeijer2002,Rietmeijer2005,Flynn2004} but slightly lower than the value of $\phi_{\rm global}\,=\,0.4$ used in the model of \citet{Skorov2012}. \citet{BabadzhanovKokhirova2009} measured the porosities of meteoroids from various meteor streams and found values for the volume filling factor between $\phi_{\rm global}\,=\,0.17$ and $\phi_{\rm global} \, = \, 1$, with most values around $\phi_{\rm global} \, = \, \sim 0.6$ (see also Table \ref{tab:metstreams}). These values are compatible with the assumption of $\phi_{\rm local} \, = \, 0.35$ for the dust aggregates in the model of \citet{Skorov2012}.
\par
It is important to mention that meteoroids associated with evolved streams (e.g., Quadrantids, Perseids, or Taurids) have a volume filling factor close to $0.6$ that is compatible with the porosities inferred by \citet{Consolmagno2008} for carbonaceous chondrites. On the other hand, the inferred properties for large aggregates are consistent with porous 81P/Wild 2 particles that, when penetrating into Stardust aerogel at $6 \, \mathrm{km \, s^{-1}}$, disrupted to produce bulbous tracks \citep{Trigo2008} and also created distinctive craters in the Al-foils \citep{Kearsley2009}. The absence of fragile meteoroids in these streams, except for perhaps relatively young October Draconids associated with comet 21P/Giacobini-Zinner, is probably due to erosive processes in the interplanetary medium as envisioned by \citet{Trigoetal2005,Trigo2013}.
\par
We conclude that the cometary meteoroids entering Earth's atmosphere are low-tensile strength bodies, with typical values for the cohesive strength of $\sim 10^4 \, \mathrm{Pa}$. However, as shown by \citet{Skorov2012} and explained above, these values are several orders of magnitude too high to explain the continued dust and gas activity of comets so that the comet meteoroids cannot be fragments from a broken homogeneous dust crust. Thus, a plausible conclusion is that comets consist of a loosely-bound ensemble of original protoplanetary dust aggregates, which are released when the comet approaches the Sun by the relatively low gas pressure of evaporating water-ice (in the inner Solar System) and other volatiles (in the outer Solar System). Their sizes of typically $1 \, \mathrm{mm}$ to $1 \, \mathrm{cm}$ give indication to the typical dust-aggregate sizes within cometary nuclei. In the next Section, we will experimentally prove that a comet nucleus formed by gravitational collapse of macroscopic dust and ice agglomerates does indeed possess an ultra-low low tensile strength to explain a continued activity as modeled by \citet{Skorov2012}.

\section{\label{sec:lab}The tensile strength of loose arrangements of dust aggregates}

\subsection{Experimental Setup}
\begin{figure*}[t!]
\centering
\includegraphics[angle=0,width=0.8\textwidth]{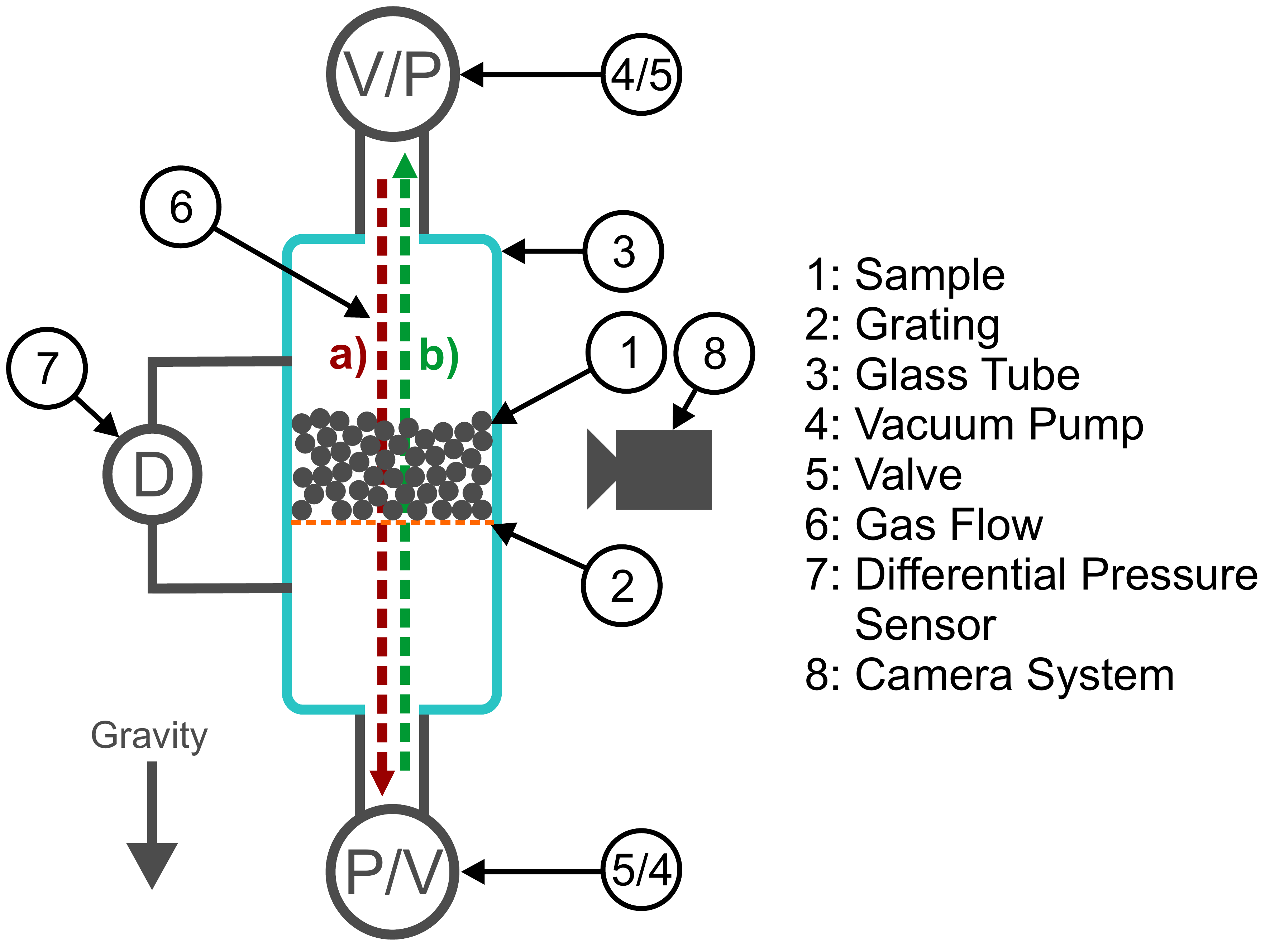}
\caption{Schematic diagram of the experimental setup used for the tensile strength measurements of granular samples. Two different versions of this experiment were used in this work. First, the gas flow was used to compress the samples (a; red arrow; the pump (P) and the valve (V) are located at the bottom and at the top of the schematic diagram, respectively). Then, the gas flow was inverted in order to measure the tensile strength of the granular samples (b; green arrow; the pump (P) and the valve (V) are located at the top and at the bottom of the schematic diagram, respectively).}
\label{fig2.1}
\end{figure*}

For the experimental simulation of the ice-free surface of a comet, we assume that the comet formed through gravitational instability and that the impact speed during the formation of the cometesimal as well as the hydrostatic pressure inside the comet nucleus were sufficiently low so that the protoplanetary dust aggregates survived intact. We measured the tensile strength of loose arrangements of dust aggregates\footnote{In this context, ``loose'' means that the aggregates were carefully poured onto the grating so that the aggregates were not compressed and the contact areas between the aggregates were not artificially increased due to the handling of the sample material.} by using the experimental setup shown in Fig. \ref{fig2.1}. The sample (1) was carefully positioned on a grating (2) inside a glass tube (3). The glass tube was then carefully and slowly evacuated by a vacuum pump (4) down to a pressure of $\sim 10\, \mathrm{Pa}$. A needle valve (5) enables the generation and control of a gas flow through the sample (6). By inversion of the gas flux direction (see Fig. \ref{fig2.1} for the two versions (a and b) of the experiment), a static compression of the samples prior to the tensile-strength measurements is feasible. The pressure difference beneath and above the sample was measured by a differential pressure sensor (7; Furness Controls FCO332). Additionally, the sample was observed by a camera system (8) during the entire duration of the experimental runs. The differential-pressure data and the recorded video stream were simultaneously stored on a computer hard disk.

\subsection{Sample Preparation, characterization, and properties}\label{Sample Preparation and characterization}
The experiments were carried out with two different kinds of sample materials. In order to validate the experimental technique (see Sect. \ref{validation}), we used black glass beads with a material density of $\rho_{\rm glass \ beads} \, = \,2100 \, \mathrm{kg \, m^{-3}}$ and five different diameters of $(40 \pm 15)\,\mathrm{\mu m}$, $(177 \pm 59)\,\mathrm{\mu m}$, $(371 \pm 126)\,\mathrm{\mu m}$, $(507 \pm 84)\,\mathrm{\mu m}$, and $(925 \pm 64)\,\mathrm{\mu m}$, respectively. Here, the errors denote the standard deviations of the diameters. The detailed size distributions of these samples can be found in \citet[Fig. 3 of][]{Gundlach2012}. The volume filling factors of a gravitationally settled packing of these samples are $0.67\pm0.02$, $0.63\pm0.02$, $0.69\pm0.02$, $0.69\pm0.02$ and $0.69\pm0.01$, respectively. The volume filling factors were estimated by measuring the occupied volume $V$ and the weight $m$ of the sample material: $\phi \, = \, \left( \,m \, / V \, \right) \, / \, \rho_{\rm glass \ beads}$.
\par
For the tensile strength measurements (see Sect. \ref{aggregates}), dust aggregates consisting of irregular-shaped, polydisperse micrometer-sized $\mathrm{SiO_2}$ particles\footnote{Manufacturer: Sigma-Aldrich.} with a material density of $\rho_{\rm SiO_2} \, = \, 2600 \, \mathrm{kg \, m^{-3}}$ were used. The size distribution of the particles was analyzed by \citet{Kothe2013} (see their Figure 3). In terms of particle number, the median size is $0.6\, \mathrm{\mu m}$, with the central 80\% of the grains falling between $\sim 0.4\, \mathrm{\mu m}$ and $\sim 1.2\, \mathrm{\mu m}$. Due to the breadth of the size distribution, the mean particle diameter in terms of mass is $2.0\, \mathrm{\mu m}$, with the central 80\% of the mass distribution falling between grain sizes of $\sim 0.8\, \mathrm{\mu m}$ and $\sim 6\, \mathrm{\mu m}$. These particle sizes are also found in meteorites and cometary IDPs so that we consider our dust sample as representative for the solar nebula. We chose $\mathrm{SiO_2}$ due to its extensive characterization in previous studies \citep[see, e.g.,][]{Blum2006a} and the fact that the specific surface energy of $\mathrm{SiO_2}$ \citep[specific surface energy: $0.02 \, \mathrm{J\,m^{-2}}$; ][see Sect. \ref{sec:application}]{BlumWurm2000} is representative of the abundant silicates in the solar nebula.
\par
\begin{figure}[t!]
\centering
\includegraphics[angle=0,width=1\columnwidth]{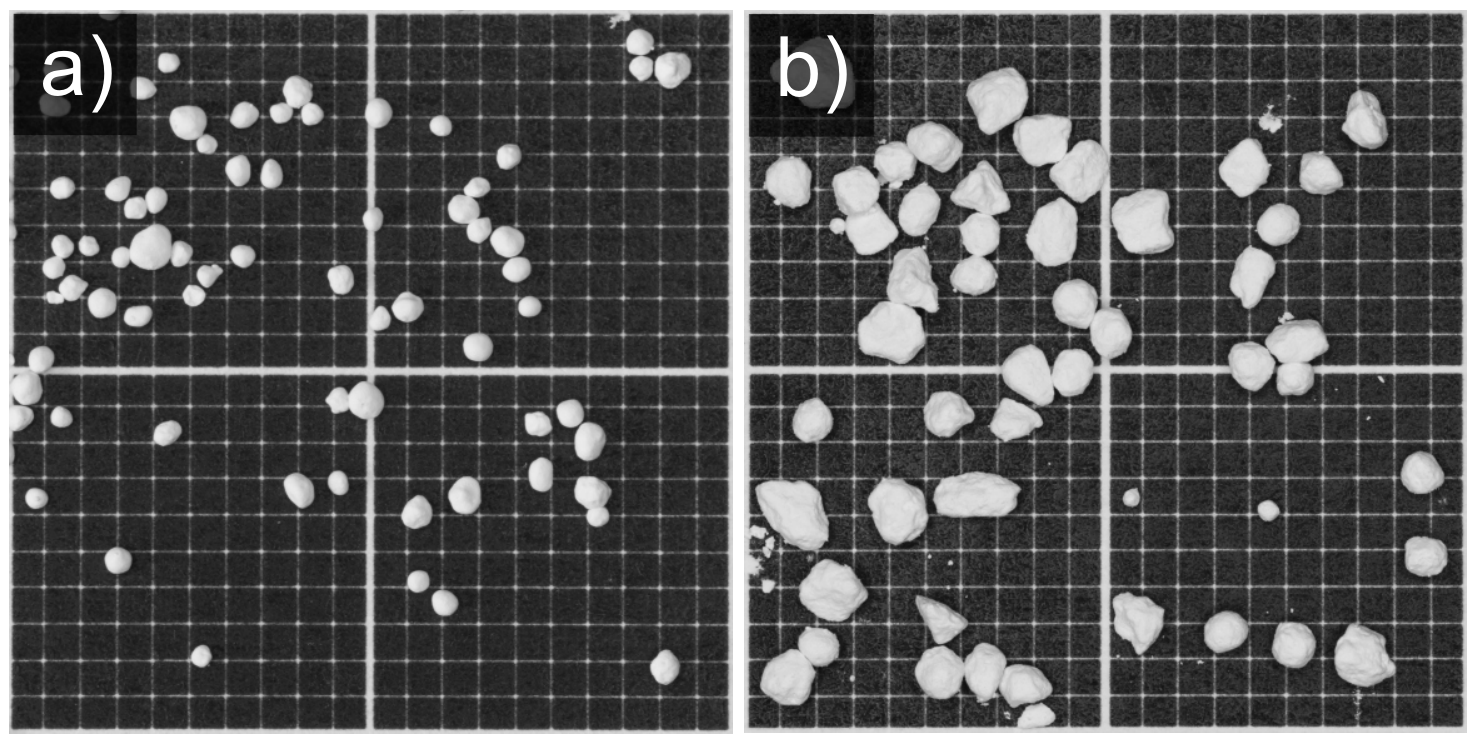}
\caption{Images of the dust aggregates DA1 (a) and DA2 (b). In order to obtain two relatively narrow size distributions (see Fig. \ref{fig2.2}), the dust aggregates were sieved before the start of the experiments.}
\label{fig2.2.1}
\end{figure}
\begin{figure}[t!]
\centering
\includegraphics[angle=180,width=1\columnwidth]{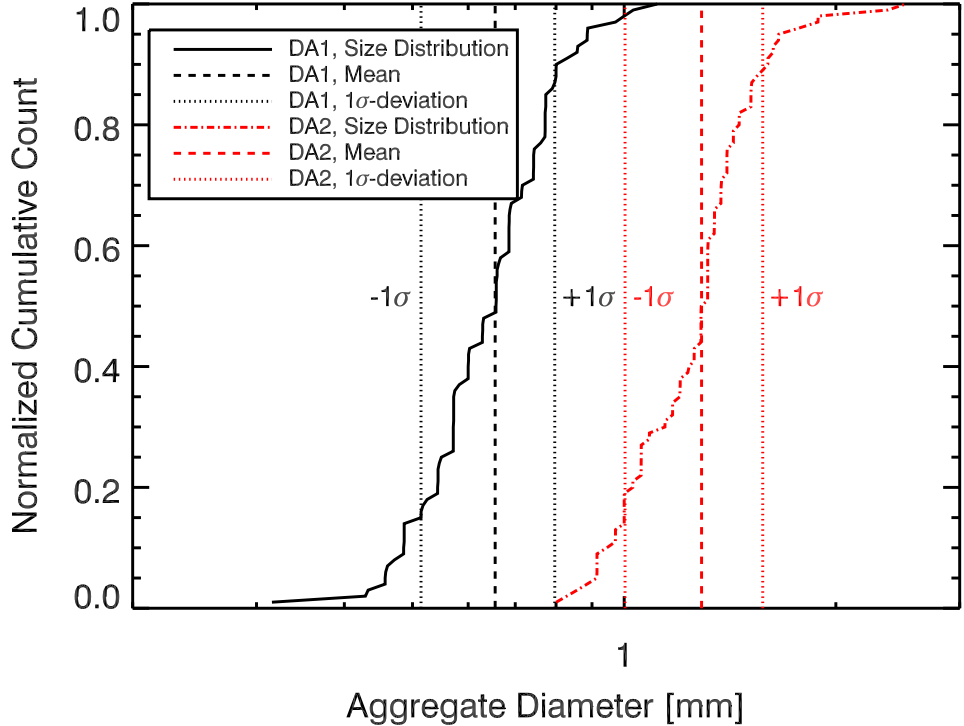}
\caption{Cumulative distribution functions of the diameters of 100 dust aggregates of types DA1 (solid black curve) and DA2 (dashed-dotted red curve). The mean diameters (dashed lines) and standard deviations (dotted lines) of the dust aggregates are $(0.66 \pm 0.14) \, \mathrm{mm}$ (DA1) and $(1.29 \pm 0.29) \, \mathrm{mm}$ (DA2), respectively.}
\label{fig2.2}
\end{figure}

The dust aggregates formed by agglomeration of the dust monomers inside the sample container. In order to obtain two samples with different sizes, we sieved the dust aggregates by using three different mesh sizes ($0.5\, \mathrm{mm}$, $1.0\, \mathrm{mm}$, and $1.6\, \mathrm{mm}$, respectively). This technique resulted in two relatively narrow size distributions (DA1 and DA2; see Fig. \ref{fig2.2.1}). The size-distribution functions of the dust aggregates were obtained by measuring the diameters of 100 individual dust aggregates for each size distribution (DA1 and DA2) by analyzing images of the dust aggregates with a computer software. Fig. \ref{fig2.2} shows the obtained diameter distributions DA1 (solid black curve) and DA2 (dashed-dotted red curve). The mean diameters (dashed lines) and standard deviations (dotted lines) of the size distributions are $(0.66 \pm 0.14) \, \mathrm{mm}$ (DA1) and $(1.29 \pm 0.29) \, \mathrm{mm}$ (DA2), respectively. The so-produced dust aggregates possess a volume filling factor of $\phi_{\rm local} = 0.35$, according to \citet{Weidling2011}. This implies a density of the dust aggregates of $\rho_{\mathrm{agg}} = \phi_{\rm local} \times \rho_{\rm SiO_2} = 910 \, \mathrm{ kg \, m^{-3}}$. Laboratory experiments have shown that these dust aggregates possess an internal tensile strength of $p_{\rm tensile\ (internal)} = 2400 \, \mathrm{Pa}$ \citep[measured for a volume filling factor of $\phi_{\rm local} = 0.41$;][]{Blum2006a} and a bulk modulus of $p_{\rm bulk\ modulus} = 25 \, \mathrm{MPa}$ \citep[derived for a density of $\rho_{\mathrm{agg}} = 910 \, \mathrm{ kg \, m^{-3}}$;][]{Guettleretal2009}.
\par
At the beginning of an experimental sequence, the sieved dust aggregates were carefully poured into the glass tube where they formed a granular packing of $25 \, \mathrm{mm}$ diameter and $17.2\mathrm{mm}$ to $25.5\mathrm{mm}$ height on top of the grating (see Fig. \ref{fig2.1}). The volume filling factor of the packing of the dust aggregates is expected to be close to the value of random close packing, $\phi_{\rm global} \approx 0.64$. Thus, the total volume filling factor of the granular sample is given by $\phi_{\rm total} = \phi_{\rm local} \, \phi_{\rm global} = 0.35 \times 0.64 = 0.22$ (see Table \ref{tab:comprop}). In order to avoid compaction of the aggregates by pouring the sample material into the glass tube, the pouring height was minimized. Due to technical reasons, the pouring height could not be reduced below $\sim 7 \mathrm{cm}$, which resulted in an impact velocity of $1.2 \, \mathrm{m \, s^{-1}}$, close to the escape velocity of a km-sized planetesimal. Thus, the experimental conditions in our experiment match those of the ice-free outer layers of comet nuclei formed by gravitational instability.

\subsection{Experimental Procedure}\label{Experimental Procedures}
In the following, we describe the different steps of our experimental procedure. All experiments were performed using this step-by-step procedure:
\begin{enumerate}
\item First, the samples (black glass beads for the validation experiments, dust aggregates for the tensile strength measurements) were poured onto the grating inside the glass tube (see Fig. \ref{fig2.1}) and the height of samples was measured.
\item Then, the glass tube was slowly evacuated until a pressure of $\sim 10\, \mathrm{Pa}$ was reached. Care was taken during the evacuation process to not destroy the samples.
\item After reaching the desired pressure, the valve in the configuration shown in Fig. \ref{fig2.1} (version (a) of the experiment) was gently opened to apply a rarefied gas flow to the sample, which statically compressed the sample for a duration of $2 \, \mathrm{min}$ (the rationale for this duration will be presented in Sect. \ref{aggregates} and Fig. \ref{fig3}). Pressure differences between $0 \, \mathrm{Pa}$ and $2000\, \mathrm{Pa}$ were used to compress the samples. In the validation experiments (see Sect. \ref{validation}), no compression was applied to the samples.
\item Thereafter, the gas flow was inverted so that the pressure difference acted against the gravity vector (see Fig. \ref{fig2.1}; version (b) of the experiment). Following Darcy's law, the gas flow through the samples caused a linear pressure decrease (i.e. a constant pressure gradient) inside the sample, if we assume a constant density of the sample.
\item The pressure gradient inside the sample was slowly increased until the applied pressure difference caused the sample to break. The break-up of the sample was observed with a camera to measure the break-up height of the sample. Additionally, the pressure differences between the bottom and the top of the sample were noted just before sample breakage to estimate the gravitational pressure (see Sect. \ref{validation}) and the tensile strength of the sample (see Sect. \ref{aggregates}). The uncertainty of the pressure-difference measurements was $0.25 \, \%$ of the measured value (manufacturer information). However, reading and noting the pressure difference exactly at the moment of the break-up also causes an estimated error of $1\, \mathrm{Pa}$. For the break-up height estimations, we assume an uncertainty of $2 \, \mathrm{mm}$. Taking these measurements errors into account implies that the uncertainty of the tensile strength measurements are dominated by the error of the break-up height.
\end{enumerate}

\subsection{\label{validation}Validation Experiments}
\begin{figure}[t!]
\centering
\includegraphics[angle=180,width=1\columnwidth]{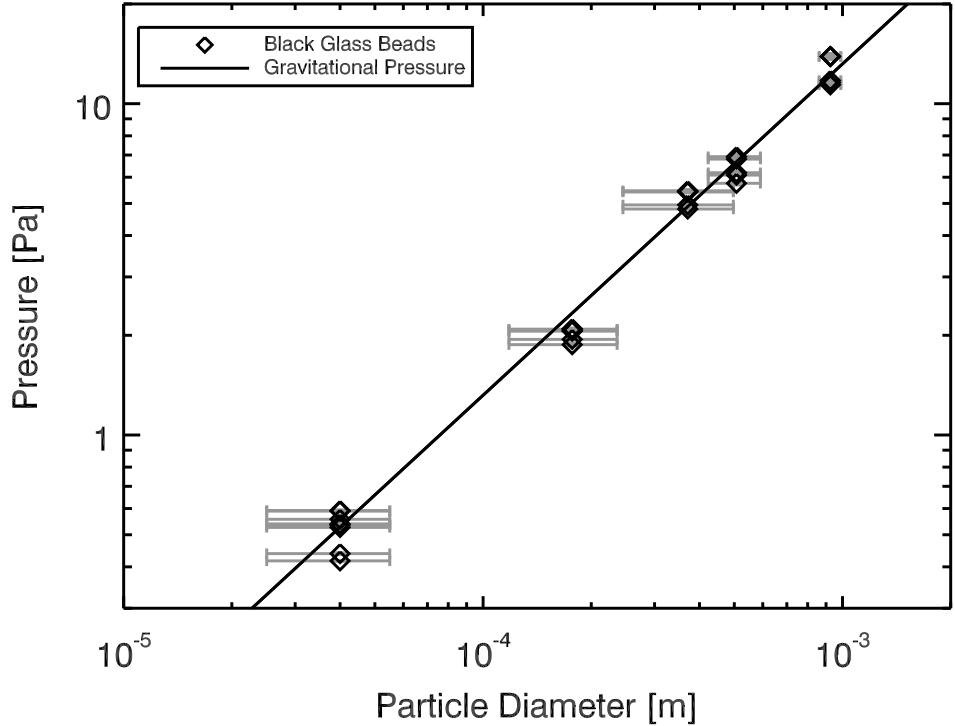}
\caption{Results of the validation experiments performed with black glass beads of different sizes ($(40 \pm 15)\,\mathrm{\mu m}$, $(177 \pm 59)\,\mathrm{\mu m}$, $(371 \pm 126)\,\mathrm{\mu m}$, $(507 \pm 84)\,\mathrm{\mu m}$ and $(925 \pm 64)\,\mathrm{\mu m}$). The diamonds show the measured cohesive strength as a function of the particle diameter. The gray error bars denote one standard deviation of the five size distributions. The black curve shows the gravitational pressure of a monolayer of the glass beads and is not a fit to the data (see Eq. \ref{eq_Basti_1}).}
\label{fig2}
\end{figure}

In order to test the feasibility of our experiment, we first used samples composed of black glass beads with five different size distributions ($(40 \pm 15)\,\mathrm{\mu m}$, $(177 \pm 59)\,\mathrm{\mu m}$, $(371 \pm 126)\,\mathrm{\mu m}$, $(507 \pm 84)\,\mathrm{\mu m}$, and $(925 \pm 64)\,\mathrm{\mu m}$; see see Sect. \ref{Sample Preparation and characterization}). In this size range, the samples have virtually no cohesion so that our method (see Sect. \ref{Experimental Procedures}) should yield the gravitational pressure of the particle layers above the break-point. No static compression was applied to the black glass bead samples before the experiments were conducted. We performed between five and eight experiments for each particle size. Breaking of the samples (i.e., the applied pressure gradient exceeds the gravitational pressure of the particle layers above the break-point) always occurred beneath the uppermost particle layer. Fig. \ref{fig2} shows the measured break-up pressures as a function of the particle size (mean particle diameter and standard deviation). The data (diamonds) follow exactly the gravitational pressure of a particle monolayer (black curve), given by
\begin{equation}
p_{\mathrm{grav}} \, = \, \phi_{\mathrm{\rm global}} \, d \, \rho_{\mathrm{glass}} \ g \, \mathrm{,}
\label{eq_Basti_1}
\end{equation}
where $\phi_{\rm global} = 0.64$ is the volume filling factor of the samples (we assume that the black glass beads have a random close packing structure inside the glass tube), $d$ is the diameter of the black glass beads, $\rho_{\rm glass} = 2100 \, \mathrm{kg \, m^{-3}}$ is the mass density of the material (see Sect. \ref{Sample Preparation and characterization}), and $g$ is the gravitational acceleration. The validation experiments prove that our method can be used to measure the weight of the particles layers per unit area and, therewith, the tensile strength of granular samples.

\subsection{\label{aggregates}Experiments with Dust Aggregates}
All measurements of the tensile strength of the dust aggregates were performed as described in Sect. \ref{Experimental Procedures}. The expected tensile strength for dust aggregates of $s = 1$ mm radii and a packing density of $\phi_{\rm global} = 0.64$ is, according to \citet{Skorov2012}, described by Eq. \ref{Eq:tsmodel} and reads $p_{\rm tensile} = 1.0$ Pa. The hydrostatic pressure of a single monolayer of dust aggregates of this size is $p_{\mathrm{grav}} = 2 \, s \,  \phi_{\mathrm{global}} \, \rho_{\mathrm{agg}}\, g = 10.7$ Pa (see Eq. \ref{eq_Basti_1}) for $\rho_{\mathrm{agg}} =910 \, \mathrm{ kg \, m^{-3}}$ ($\rho_{\mathrm{agg}} = \phi_{\rm local} \times \rho_{\rm SiO_2}$; see Sect. \ref{Sample Preparation and characterization}). As the sample typically breaks several monolayers below its upper surface, a direct measurement of the tensile strength is difficult. Thus, we increased the tensile strength by static compression of the samples with a rarefied gas flow (see Sect. \ref{Experimental Procedures} and Fig. \ref{fig2.1}; version (a) of the experiment). After the compression of the samples, the direction of the gas flow was inverted in order to break the samples. For both directions, we derived the gas-flow velocity by measuring the flow rate with a flow meter. The maximum speed of the rarefied gas flowing through the dust aggregate samples was $5.7 \, \mathrm{m\, s^{-1}}$. Thus, the dynamical gas pressure is $1.9 \times 10^{-2}\, \mathrm{Pa}$, which is negligible in comparison with the static pressure caused by the gas flow and the resistivity of the dust sample to the gas flux. Hence, we only take the static pressure of the flowing gas into account.
\par
\begin{figure}[t!]
\centering
\includegraphics[angle=0,width=1\columnwidth]{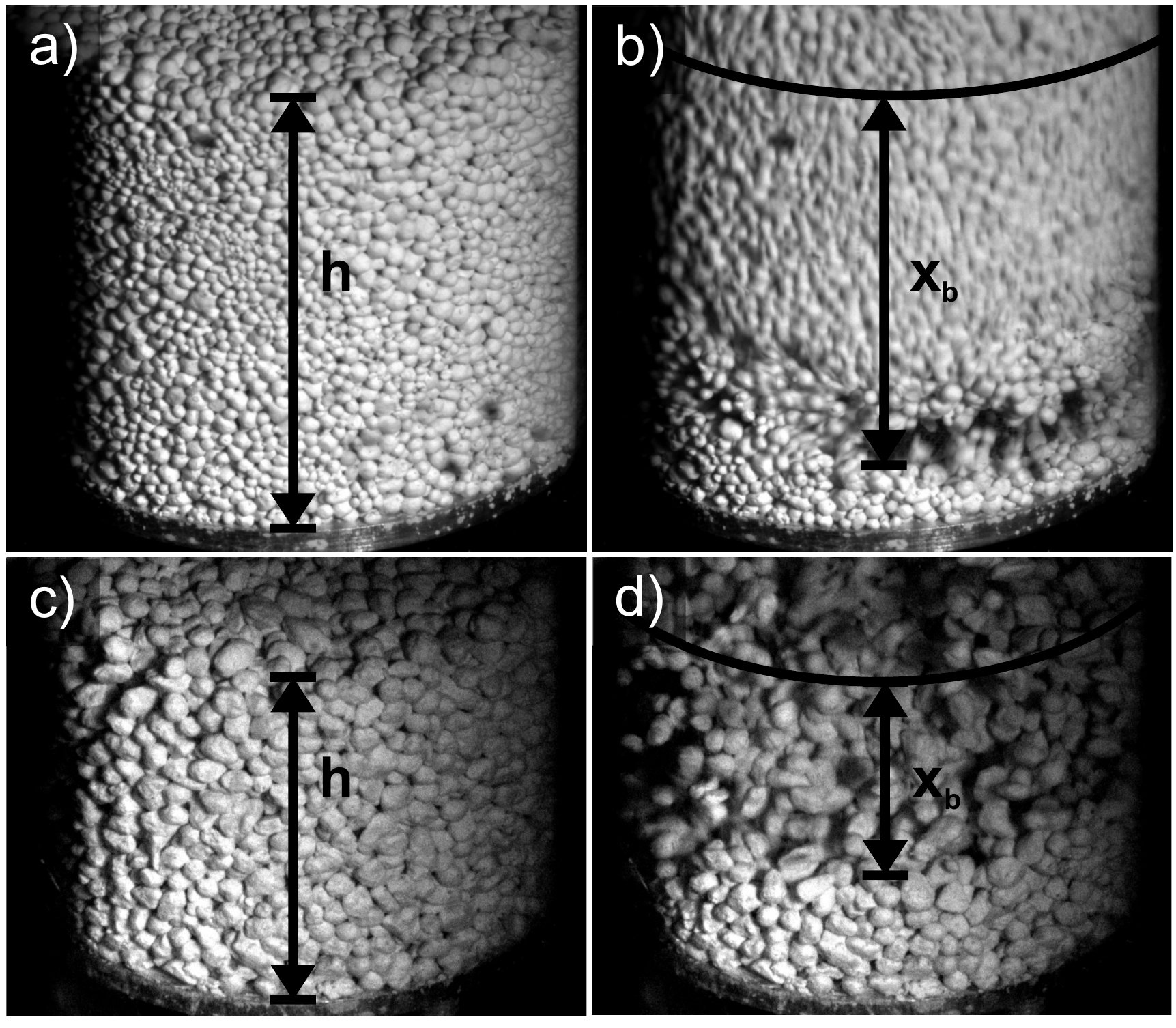}
\caption{Dust-aggregate samples DA1 (a and b) and DA2 (c and d) before (left images) and during the break-up of the samples (right images). The height of the samples was $h = 22.5 \, \mathrm{mm}$ (DA1) and $h = 17.2 \, \mathrm{mm}$ (DA2), respectively. The break-up points of the samples were estimated to be $x_b = 18.8 \, \mathrm{mm}$ (DA1) and $x_b = 10.4 \, \mathrm{mm}$ (DA2, respectively).}
\label{fig3.1}
\end{figure}

After the inversion of the gas flow direction, the gas flow through the sample was slowly increased. At relatively low gas flows (i.e., for small pressure gradients), the sample stayed intact. Above a threshold for the flow rate (or pressure gradient), the gas flow led to a break-up of the sample (see Fig. \ref{fig3.1}). In the case of the dust aggregate samples, breaking occurred away from the uppermost and lowermost layers, most often close to the middle of the sample. This observation can be explained by an inversion of the Janssen effect \citep{Janssen1895,Sperl2006}, which describes the saturation of pressure with depth in a granular material packed into a tubular volume. The saturation of pressure with depth is caused by the force propagation in granular materials, which can efficiently transport the load of the particle layers onto the side walls. Typically, the Janssen effect becomes important when the height of the granular sample exceeds the radius of the tube \citep[see Fig. 13 in][]{Sperl2006}. In our case, the normal Janssen effect is not important because the gas pressure always has to overcome the tensile strength plus the weight of the particle layers to break the sample, i.e. to lift the dust aggregates. However, to break the dust-aggregate samples, we applied a pressure gradient against gravity. In this case, the load caused by the gas pressure is also propagating through the particle layers onto the side walls. This implies that the applied pressure gradient can only effectively act on the dust-aggregates layers if the height of the covering layers does not exceed the tube radius by much. This hypothesis is in full agreement with our experiments, because the height of the samples was between $17.2\,\mathrm{mm}$ and $25.5\,\mathrm{mm}$ (thus, between 1.4 and 2.0 tube radii) and we observed that the break-up of the samples always occurred in the lower half of the samples. Thus, the Janssen effect ensures that the samples never break-up at the bottom if the sample height is larger than approximately the radius of the tube. For sample heights smaller than the radius of the tube, break-up of the samples often occurred at the bottom of the sample, i.e., between sample and grating. These experiments were not used to estimate the tensile strength of the dust-aggregate samples.
\par
\citet{SchnitzleinEisfeld2001} investigated the influence of the friction between the confining walls and the granular material by comparing the hydraulic radius theory with experimental data. They found that the influence of the confining walls becomes important for $D\, / \, d \, < \, 10$, where $D$ and $d$ are the tube and particle diameters, respectively. Thus, we calculated the tube to particle diameter ratio for our samples: $D\, / \, d \, = \, 37.8$ (DA1 samples) and $D\, / \, d \, = \, 19.4$ (DA2 samples). Thus, we conclude that in our experiments the friction between the confining walls an the sample material are not affecting the pressure gradient in the samples.
\par
From the measured pressure difference between the bottom and the top of the samples just before break-up, $\Delta p = p_{\mathrm{bottom}} \, - \, p_{\mathrm{top}}$, and from the break-up height estimations, $x_b$, the pressure difference between the break-point and the surface of the sample,
\begin{equation}
\Delta p(x_b) \, = \, \Delta p \, \frac{x_b}{h} \, \mathrm{,}
\label{eq_Basti_3}
\end{equation}
can be derived, assuming a linear pressure drop in the sample \citep{Carman1937}. The break-up height was estimated by analyzing the video data. The images before and after the break-up of the samples (see, e.g., Figs. \ref{fig3.1}.a and \ref{fig3.1}.b) were compared in order to determine the distance between the top of the sample and the position at which the sample material was lifted (break-up height). From Eq. \ref{eq_Basti_3}, the tensile strength of the dust-aggregate samples can be calculated by
\begin{equation}
p_{\mathrm{tensile}} \, = \, \Delta p(x_b) \, - \, p_{\mathrm{grav}}(x_b)\, \mathrm{.}
\label{eq_Basti_2}
\end{equation}
The second term of the right hand side of equation \ref{eq_Basti_2} is the gravitational pressure at the break-point $x_b$, caused by the weight of the dust-aggregate layers above and reads
\begin{equation}
p_{\mathrm{grav}}(x_b) \, = \, \rho_{\mathrm{total}} \, g \, x_b  \, \mathrm{.}
\label{eq_Basti_4}
\end{equation}
Here, $\rho_{total} \, = \, \rho_{\mathrm{SiO_2}} \, \phi_{\mathrm{global}} = \, 0.572 \, \mathrm{kg \, m^{-3}}$ (see Sect. \ref{Sample Preparation and characterization}) is the density of the dust-aggregate sample.
\par
\begin{figure}[t!]
\centering
\includegraphics[angle=180,width=1\columnwidth]{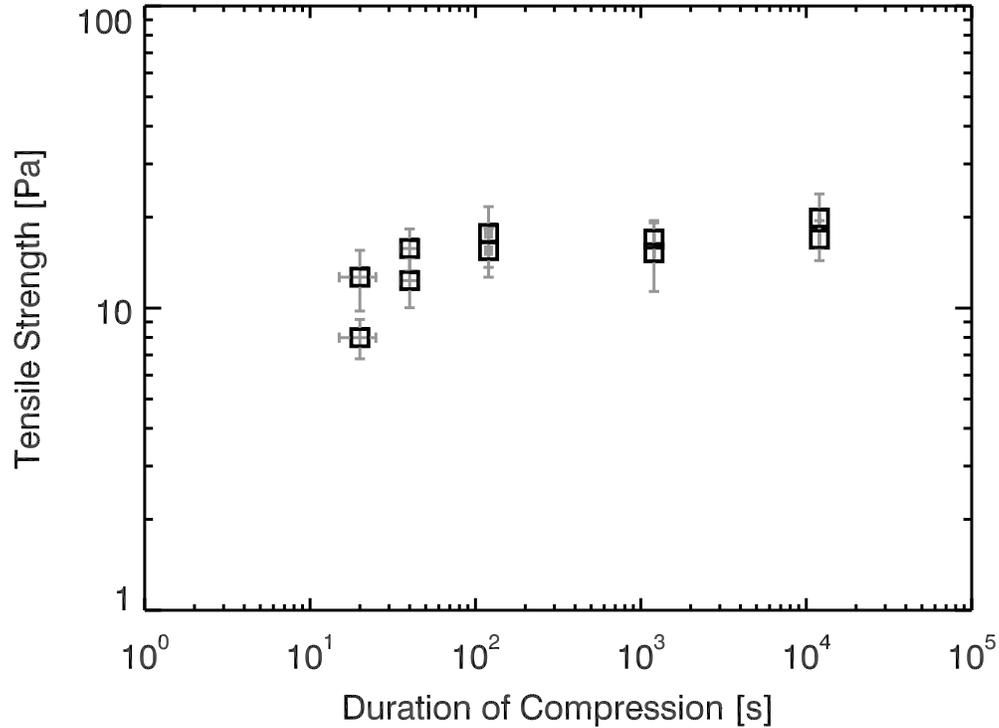}
\caption{Tensile strengths of the dust-aggregate samples (DA1) as a function of the duration of the static compression. In this case, a constant compression of $\Delta p = 1000 \, \mathrm{Pa}$ was applied. The height of the dust-aggregate samples was $h \approx 18 \, \mathrm{mm}$.}
\label{fig3}
\end{figure}

As the compression dynamics of the dust-aggregate samples were unknown, we first investigated the influence of the duration of compression on the resulting tensile strength. Therefore, the samples were statically compressed with a constant pressure difference of $\Delta p = 1000 \, \mathrm{Pa}$ and the exposure time to the compression was varied between $20 \, \mathrm{s}$ and $12,000 \, \mathrm{s}$. Since the compressive pressure could not be instantaneously switched on, an uncertainty of the pressure exposure of $5\,\mathrm{s}$ was assumed. After the compression, the tensile strengths of the compressed samples were measured. Fig. \ref{fig3} shows the derived tensile strength of the dust aggregates for the different durations of the compression, following equation \ref{eq_Basti_2}. The tensile strength increases slightly with the duration of the compression for short exposure times and reaches a constant value for exposures of $\sim 120 \, \mathrm{s}$ or above. Thus, for the determination of the tensile strength, we compressed all samples for $120 \, \mathrm{s}$ with pressure differences between $0 \, \mathrm{Pa}$ and $2000\, \mathrm{Pa}$ and then performed the tensile-strength measurements.
\par
\begin{figure}[t!]
\centering
\includegraphics[angle=180,width=1\columnwidth]{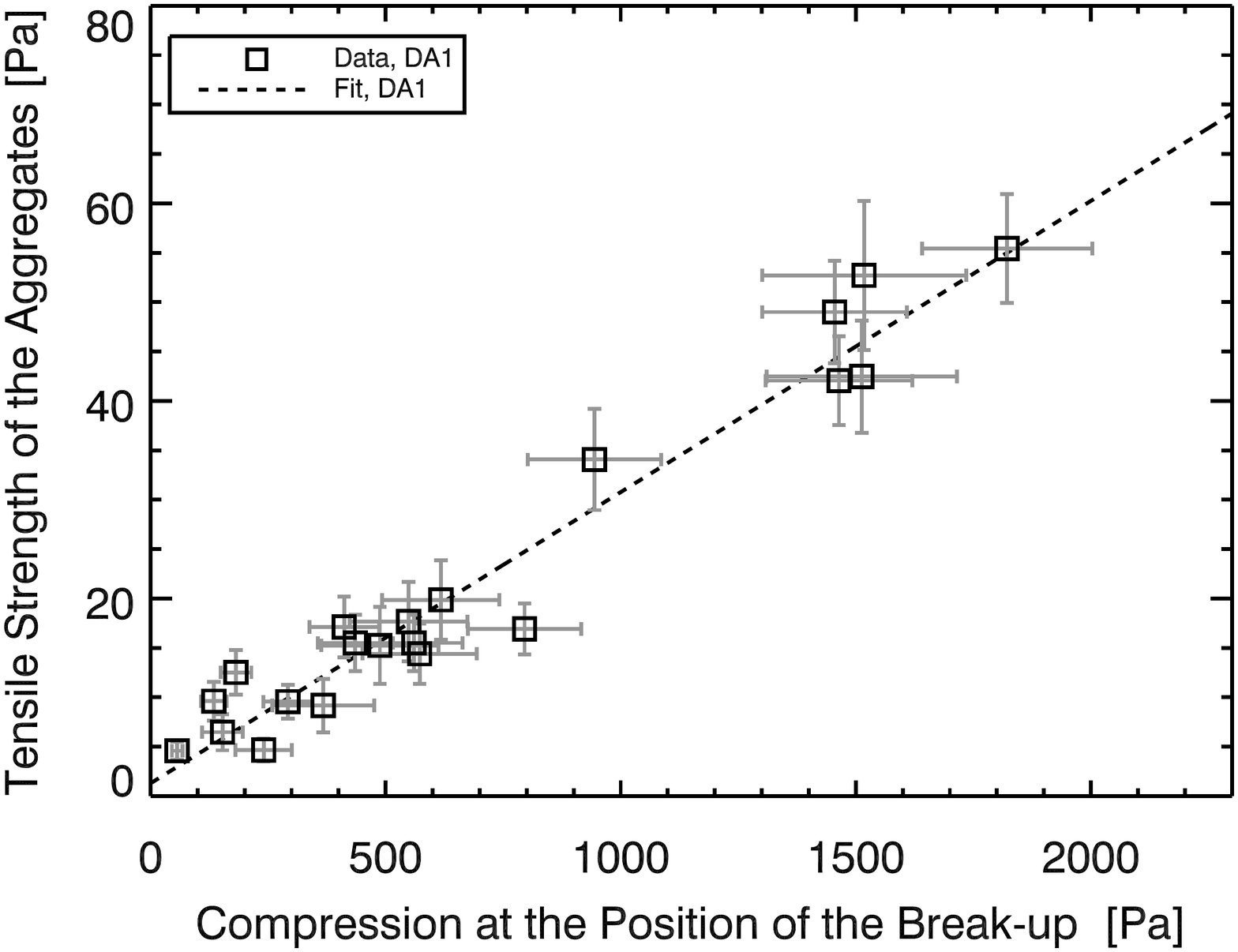}
\caption{Derived tensile strengths (see Eqs. \ref{eq_Basti_3} - \ref{eq_Basti_4}) of the DA1 samples as a function of the compression pressure at the position of the break-up (squares). The dashed line shows the least-squares linear fit applied to the data. The uncertainties of the tensile strength measurements are denoted by the gray error bars. The uncompressed tensile strength of the DA1 samples is $(1.3 \pm 0.9) \, \mathrm {Pa}$. The slope of the tensile strength vs compression curve is $(2.9 \pm 0.2 ) \times 10^{-2}$.}
\label{tensile1}
\end{figure}
\begin{figure}[t!]
\centering
\includegraphics[angle=180,width=1\columnwidth]{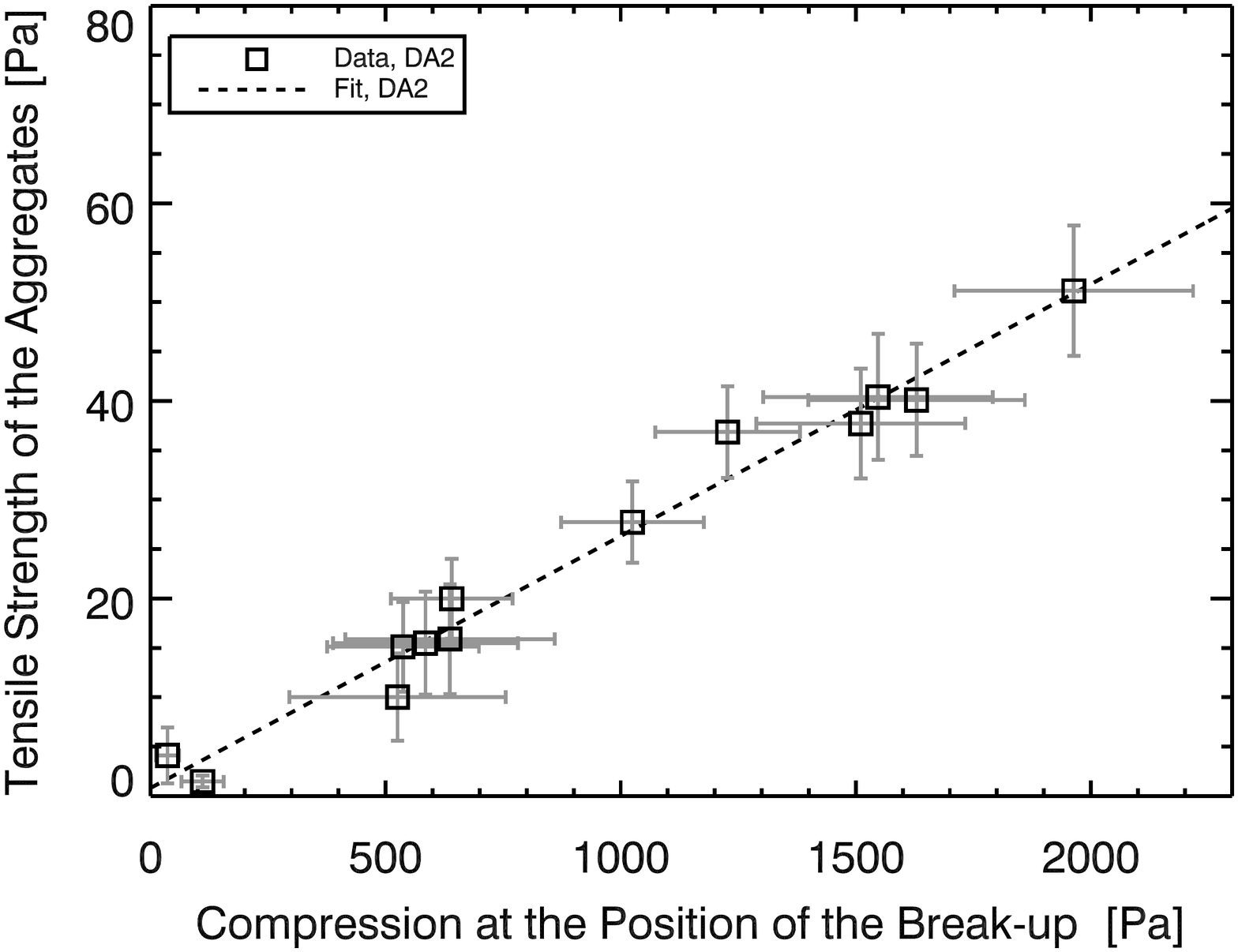}
\caption{Derived tensile strengths (see Eqs. \ref{eq_Basti_3} - \ref{eq_Basti_4}) of the DA2 samples as a function of the compression pressure at the position of the break-up (squares). The dashed line shows the least-squares linear fit applied to the data. The uncertainties of the tensile strength estimations are denoted by the gray error bars. The uncompressed tensile strength of the DA2 samples is  $(0.8 \pm 0.7) \, \mathrm {Pa}$. The slope of the tensile strength vs compression curve is $(2.6 \pm 0.1 ) \times 10^{-2}$.}
\label{tensile2}
\end{figure}

In total, we performed 22 experimental runs with the samples of type DA1 and 13 experiments with DA2 samples. Figs. \ref{tensile1} and \ref{tensile2} show the derived tensile strengths of the samples, following Eq. \ref{eq_Basti_2}, as a function of the applied compression. Both samples show a linear increase of the tensile strength with increasing compression. Obviously, the dust-aggregate samples remember the strength of the compression applied before. We fit a linear curve to the data in order to describe the increase of the tensile strength with increasing compression at the position of the break and found a slope of $(2.9 \pm 0.2)\times 10^{-2}$ for the DA1 samples and $(2.6 \pm 0.1)\times 10^{-2}$ for the DA2 samples. The effect of strengthening of granular powders by applying a static pressure to the samples has been observed before \citep[but at much higher stresses; see Fig 5 in][]{Tomas2004} but never for dust aggregates and at such a low degree. The experiments performed by \citet{Tomas2004} have also shown a linear correlation between the tensile strength of the samples and the applied compression.
\par
To derive the uncompressed tensile strengths of the samples, we extrapolated the fit curves to $\Delta p = 0 \, \mathrm{Pa}$. Using fixed slopes of $2.9 \times 10^{-2}$ for the DA1 samples and $2.6 \times 10^{-2}$ for the DA2 samples, we derived uncompressed tensile strengths of $(1.3 \pm 0.9) \, \mathrm {Pa}$ (DA1) and $(0.8 \pm 0.7) \, \mathrm {Pa}$ (DA2), respectively. In comparison to homogeneous $\mathrm{SiO_2}$ dust samples \citep[$p_{\rm tensile} \approx 1 \, \mathrm{kPa}$;][]{Blum2006a}, our dust-aggregate samples are orders of magnitude weaker, due to the small number of adhesive contacts.
\par
The derived uncompressed tensile strengths of the dust-aggregate samples are in  good agreement with the model of \citep{Skorov2012}. Referring to Eq. \ref{Eq:tsmodel}, the model predicts a tensile strength of $1.4 \, \mathrm {Pa}$ and $0.9 \, \mathrm {Pa}$ for dust aggregates with radii of $0.66 \, \mathrm{mm}$ (mean radius of the DA1 samples) and $1.29 \, \mathrm{mm}$ (mean radius of the DA2 samples), respectively. For the calculations, we used a volume filling factor of $\phi_{\rm global} = 0.64$ (random close packing of the dust-aggregate samples; see Sect. \ref{Sample Preparation and characterization}). The comparison between the model and the experiments also confirms the systematic decrease of the tensile strength with increasing dust-aggregate size as predicted by the model.

\section{\label{sec:application}Application to Comet Nuclei}
\subsection{Global cometary activity}
\begin{figure}[t!]
\centering
\includegraphics[angle=180,width=1\columnwidth]{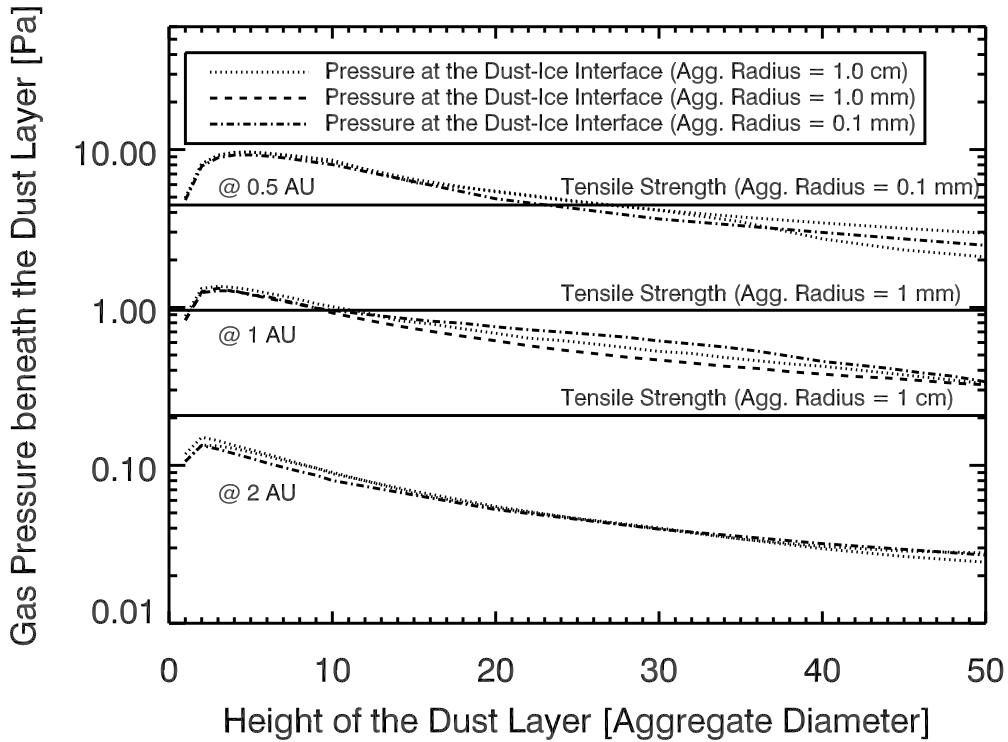}
\caption{\label{cometarysurface}Comparison between gas pressures at the dust-ice interface (curves) and the tensile strengths of the dust-aggregate layers (lines), the latter computed by using the model of \citet{Skorov2012} as a function of the height of the covering dust-aggregate layers, in units of the diameter of the aggregates. The three curves and lines denote dust-aggregate radii of $10 \, \mathrm{mm}$ (dotted), $1 \, \mathrm{mm}$ (dashed) and $0.1 \, \mathrm{mm}$ (dash-dotted). The pressure at the dust-ice interface was determined by computing the temperature of the dust-covered ice surface by numerically solving the heat transfer equation and by taking the permeability of the dust-aggregate layer into account. The calculations were performed for heliocentric distances of $0.5 \ \mathrm{AU}$ (upper curves), $1 \ \mathrm{AU}$ (middle curves), and $2 \ \mathrm{AU}$ (lower curves).}
\end{figure}

Visible cometary activity is due to the release of significant amounts of dust from the cometary surface, which can then be detected by scattering of solar radiation. The production of dust can only be sustained over an extended period of time if the thickness of the volatile-free dust layer of the comet nucleus is not too high. \citet{Skorov2012} suggested that (locally) the dust layer grows in thickness when it is too thin, because the evaporating ice molecules can escape almost unimpeded so that the pressure at the bottom of the dust layer is insufficient to release the dust. With growing thickness of the dust layer, the amount of heat that reaches the ice surface is reduced, which decreases the sublimation rate, but the gas permittivity of the dust layer is increased, which leads to an increased static gas pressure under the dust layer. Due to the combination of both effects, the pressure difference between bottom and top of the dust layer becomes maximal at a certain thickness of the dust layer (see Fig. \ref{cometarysurface} and the discussion below) and decreases again when the dust layer becomes thicker. Dust activity of comet nuclei is only possible if the pressure drop over the dust layer exceeds its tensile strength (plus its gravitational strength, but the latter is normally negligibly small for comets). In this case, the whole dust layer is released and the process cycle starts again.
\par
Homogeneous dust layers consisting of micrometer-sized siliceous materials exhibit a low tensile strength on the order of $p_{\rm tensile} \sim 10^3-10^4 \, \mathrm{Pa}$, as shown experimentally by \citep{Blum2006a}, which has been confirmed by the remote studies of decaying comets \citep[see, e.g.,][]{Blum2006a} and the compilation of the typical dynamic strengths of cometary meteoroids (see Table \ref{tab:metstreams} in Sect. \ref{sec:comtensile}). Dust activity against such tensile strengths requires water-ice sublimation temperatures and outgassing rates way too high to be realistic. If, however, according to the model by \citet{Skorov2012}, the near-surface regions of comet nuclei consist of loosely-bound dust aggregates, the expected tensile strengths of the ice-free dust-aggregate layers are ultra-low and can be exceeded by the gas pressure of the evaporating volatiles beneath. For such a geometry, the model by \citet{Skorov2012} predicts tensile strengths according to Eq. \ref{Eq:tsmodel} (see Sect. \ref{sec:introduction}). A comparison with our experimental results shows a quantitative agreement so that we can conclude that Eq. \ref{Eq:tsmodel} describes the correct van der Waals tensile strength of ice-free dust-aggregate layers.
\par
It is important to mention that this ejection mechanism of cometary dust layers is able to simultaneously release a large number of dust aggregates that directly can be driven away due to the quick sublimation of fresh ices under solar irradiation. It can, thus, also be a contribution to explain cometary outbursts \citep{Sekanina1982,Trigoetal2010}.
\par
We calculated the gas pressure of water ice covered by dust-aggregate layers of different thicknesses (see Fig. \ref{cometarysurface}). The gas pressure at the dust-ice interface was determined by computing the temperature profile inside the dust-aggregate layer and inside the ice underneath. For the computations, we used the heat conductivity model of dust-aggregate layers in vacuum \citep[see Sect. 5.2 in][]{Gundlach2012}. Furthermore, we assumed solid (i.e., non-porous), hexagonal water ice beneath the dust-aggregate layer. From the temperature profile, the temperature of the ice surface was derived under the assumption that the energy transported to the ice surface is totally consumed by the sublimation process (i.e., no energy is transported through the ice into deeper layers of the nucleus). Thus, the temperature of the ice interface determines the sublimation pressure of the ice for the different thicknesses of the dust-aggregate layers. We then used the gas permittivity of the dust-aggregate layer \citep[see Eq. 19 in][]{Gundlach2011} to calculate the static gas pressure at the ice-dust interface. For the calculations, we used a bond albedo of 0.04, which is a typical value for cometary surfaces \citep[see, e.g.,][]{Keller1988,Brownlee2004,AHearn2005,Groussin2007,AHearnetal2011}.
\par
Fig. \ref{cometarysurface} shows the resulting static gas pressures at heliocentric distances of $0.5 \ \mathrm{AU}$ (upper curves), $1 \ \mathrm{AU}$ (middle curves), and $2 \ \mathrm{AU}$ (lower curves) for different thicknesses of the dust-aggregate layer and three different dust-aggregate radii of $10 \, \mathrm{mm}$ (dotted curves in Fig. \ref{cometarysurface}), $1 \, \mathrm{mm}$ (dashed curves) and $0.1 \, \mathrm{mm}$ (dash-dotted curves). With increasing distance of the comet to the Sun, the total amount of received and, thus, absorbed energy by the surface layer decreases. This implies a decrease of the static gas pressure for increasing heliocentric distance. Additionally, the tensile strengths of the three different dust-aggregate layers, following the model by \citet{Skorov2012} are shown by the respective horizontal lines. In the case of cm-sized dust aggregates, the static gas pressure beneath the dust layers, caused by the outgassing of water ice, exceeds the tensile strength of the dust-aggregate layer for thicknesses between two and more than 50 dust-aggregate diameters. For mm-sized dust aggregates, this range shrinks to between two and about ten monolayers. Thus, lifting of dust aggregates with a diameter larger than $0.6 \, \mathrm{mm}$ from a cometary surface at $1 \, \mathrm{AU}$ can be explained by the outgassing of water ice if the thickness of the dust layer is smaller than $\sim 10$ aggregate diameters. However, with decreasing aggregate size, the tensile strength of the dust-aggregate layer increases (see Eq. \ref{Eq:tsmodel}). Thus, below a critical dust-aggregate size, the static gas pressure is not sufficient to overcome the tensile strength of the dust-aggregate layer and the sublimation of water ice cannot destroy the dust layer (see, e.g., the results for the $0.1 \, \mathrm{mm}$ dust aggregates in Fig. \ref{cometarysurface}).
\par
\begin{figure}[t!]
\centering
\includegraphics[angle=0,width=0.9\columnwidth]{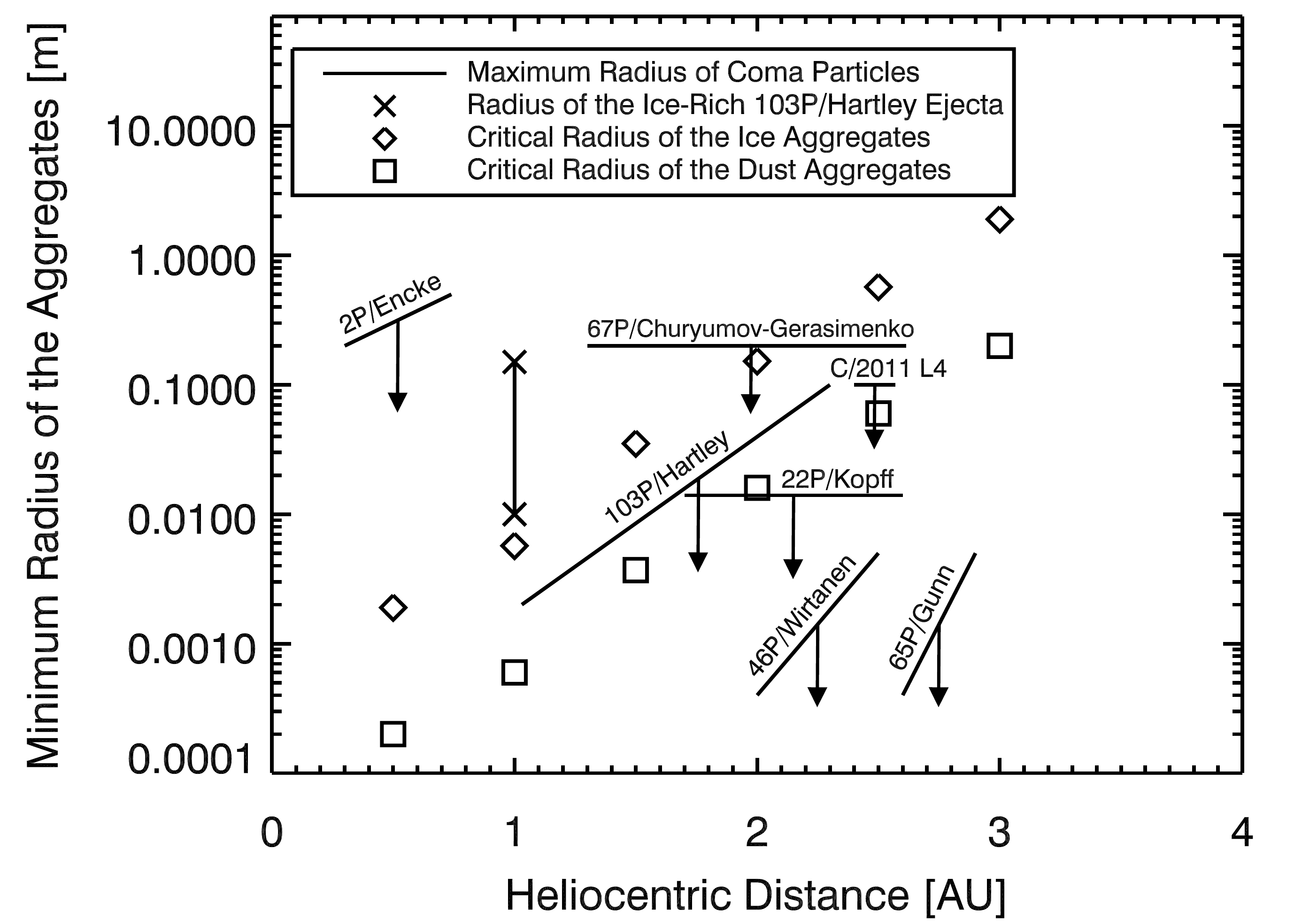}
\caption{Minimum radius of dust aggregates (squares) and ice aggregates (diamonds) that can be released by water-ice sublimation as a function of the heliocentric distance. Layers composed of aggregates larger than these sizes are released from the comet nucleus by the static pressure caused by water-ice sublimation (i.e., the static gas pressure exceeds the tensile strength of the aggregate layer). Surface layers composed of smaller aggregates possess tensile strengths higher than the static gas pressure and are, thus, not destroyed by the sublimation of water ice. For comparison, the estimated sizes of the ejected ice-rich particles from the nucleus of comet 103P/Hartley during the flyby of the EPOXI spacecraft \citep{AHearnetal2011} are shown by the crosses. Additionally, the solid lines represent the maximum radii of coma particles derived from observations of the following comets: 2P/Encke \citep{Epifani2001}, 103P/Hartley \citep{Epifani2001}, 67P/Churyumov-Gerasimenko \citep{Moreno2004}, 22P/Kopff \citep{Moreno2012}, 46P/Wirtanen \citep{Colangeli1998}, 65P/Gunn \citep{Colangeli1998}, and P/2012 T1 \citep[Panstarrs;][]{Moreno2013}. The downward arrows indicate that only the maximum radius of the derived size distributions are presented.}
\label{min_size}
\end{figure}

It is obvious that a critical dust-aggregate size exists above which activity is possible and below which the comet surface is inactive. We derived the critical radius of the dust aggregate for which the maximum static gas pressure equals the tensile strength of the dust-aggregate layer. Hence, dust-aggregate layers composed of larger dust clumps are destroyed by the static pressure. The tensile strength of dust-aggregate layers composed of clumps smaller than the critical size is higher than the static gas pressure caused by the water ice sublimation. In this case, the sublimation of water ice is not sufficient to blow off the dust-aggregate layer. Fig. \ref{min_size} shows the calculated critical radii of dust aggregate for different heliocentric distances between 0.5 and 3.0 AU (squares). The calculations were also performed for pure water-ice-aggregate layers and the results are additionally presented in Fig. \ref{min_size} (diamonds). In the case of the ice-aggregate layers, the proportionality between the tensile strength and the specific surface energy, $p_{\rm tensile} \propto \gamma$ \citep[see][for details]{Weidling2011}, was taken into account in order to derive the tensile strength of the ice-aggregate layers. The specific surface energy of water ice \citep[$\gamma_{\rm ice} = 0.19 \, \mathrm{J\,m^{-2}}$;][]{Gundlach2011b} is approximately ten times higher than the specific surface energy of dust \citep[$\mathrm{SiO_2}$; $\gamma_{\rm dust} = 0.02 \, \mathrm{J\,m^{-2}}$;][]{BlumWurm2000}, which implies that the tensile strength of the ice-aggregate layer is roughly a magnitude higher than the tensile strength of the dust-aggregate layer. This means that at the same heliocentric distance, only bigger aggregates can be lifted from the ice-aggregate layer in comparison to the dust-aggregate layer (i.e., the critical radii of the ice aggregates is always bigger than the critical radii of the dust aggregates in Fig. \ref{min_size}). With decreasing heliocentric distance, the rate of sublimation increases rapidly. This implies that ever smaller dust aggregates can be released from the cometary surface when the comet approaches the Sun. Thus, the onset of water driven dust activity can be used to infer the dust-aggregate size ejected by the outgassing from the cometary nucleus. Fig. \ref{min_size} shows that mm- to cm-sized dust particles can be released due to water ice activity at heliocentric distances between $\sim 1 \, \mathrm{AU}$ and $\sim 2\mathrm{AU}$. This result is in a good agreement with the observations of cometary dust (see Sect. \ref{sec:comcomp}). Additionally, the maximum radii of coma particles derived from observations of the following comets are shown in Fig. \ref{min_size} (solid lines with downward arrows): 2P/Encke \citep{Epifani2001}, 103P/Hartley \citep{Epifani2001}, 67P/Churyumov-Gerasimenko \citep{Moreno2004}, 22P/Kopff \citep{Moreno2012}, 46P/Wirtanen \citep{Colangeli1998}, 65P/Gunn \citep{Colangeli1998}, and C/2011 L4 \citep[Panstarrs;]{Moreno2013}. Except for the rather small maximum dust sizes observed for comets 46P/Wirtanen and 65P/Gunn, all observational data are in agreement with the model. The release of relatively small particles from the two exceptional comets at heliocentric distances larger than $\sim 2 \, \mathrm{AU}$ cannot be explained by the model and is either caused by a fast fragmentation of the dust aggregates within the coma or by the outgassing of other, more volatile materials than water ice.
\par
In addition, hydrostatic consolidation stresses inside the comet nucleus can also lead to an increase of the tensile strength of comets. The hydrostatic pressure inside a spherical body of constant density $\rho$ reads
\begin{align}\label{eq:hydpress}
  P(r) \,  &= \, \frac{2}{3} \, \pi \, \rho^2 \, G \, \left[ \, R^2 \, - \, r^2 \right] \\
           &= \, 14 \, {\rm Pa} \,\times \left(\frac{\rho}{1000 \, {\rm kg \, m^{-3}}}\right)^2 \times \, \left[ \left( \frac{R}{\rm 1 \, km}\right)^2 \, - \, \left(\frac{r}{\rm 1 \, km}\right)^2 \right] \,\\
           & \approx \, 28 \, {\rm Pa} \,\times \left(\frac{\rho}{1000 \, {\rm kg \, m^{-3}}}\right)^2 \times \, \left( \frac{R}{\rm 1 \, km}\right) \, \times \, \left(\frac{d}{\rm 1 \, km}\right)  \,
\end{align}
with $G$, $R$, $r$, and $d$ being the gravitational constant, the overall radius of the body, the radius at which the pressure is to be determined, and the corresponding depth under the surface, respectively, with $d\,=\,R-r$. Here, $r\, = \,0$ and $d\,=\,R$ mean the center of the body and $r\,=\,R$ and $d\,=\,0$ describe its surface. For a comet with $R\, =\, 10 \, \mathrm{km}$, we get $P\, = \,265 \, \mathrm{Pa}$ consolidation stress for $d\,=\,1\, \mathrm{km}$ below its surface. According to Figs. \ref{tensile1} and \ref{tensile2}, such a compressive pressure leads to an increase in the tensile strength of $p_{\mathrm{tensile}} \sim 8 \, \mathrm{Pa}$. Although the effect is not large and requires relatively large comets and depths, it can can lead to a shift in activity over time. Consequently, old comets (for which today's surface was originally deeper inside the comet nucleus) are expected to be less active than new comets, if there is no re-organization of the dust aggregates after the ice has evaporated. This can, in principle, lead to the extinction of comets.

\subsection{Localized cometary activity}
Spacecraft flybys (Giotto\footnote{http://sci.esa.int/science-e/www/area/index.cfm?fareaid=15}, \citealp{Schwehm2006}; Deep Space 1\footnote{http://www.jpl.nasa.gov/missions/details.php?id=5870}, \citealp{Rayman2001}; Stardust\footnote{http://stardust.jpl.nasa.gov/home/index.html}, \citealp{Brownlee2003}; Deep Impact\footnote{http://www.nasa.gov/mission$\_$pages/deepimpact/main/index.html}, \citealp{Blume2005}; EPOXI\footnote{http://www.nasa.gov/mission$\_$pages/epoxi/index.html}, \citealp{AHearnetal2011}) in the past have shown that the dust emission is not homogeneously distributed over the comet-nucleus surface. Localized emission of dust (jets) was observed for all cometary nuclei visited by spacecraft missions \citep{Keller1987,Thomas2009}. In the case of comet 9P/Tempel, the only comet consecutively visited by the Deep Impact and the Stardust-NExt mission, the localized activity has led to observable changes of the surface morphology \citep{Veverka2013}. Maintaining localized dust activity can be explained by our model (see Fig. \ref{cometarysurface}), because active regions (water driven dust activity) are able to periodically and completely remove the dust when the gas pressure exceeds the tensile strength of the covering dust layer (parts of the curves left of the maxima in Fig. \ref{cometarysurface}). This naturally explains the observations by \citep{Sunshine2006} that patches of ice exist on the surface of Comet 9P/Tempel. After the removal of the covering dust layer down to the ice-dust boundary, the thickness of the dust layer starts to grow again, due to the evaporation of the icy material until the gas pressure again reaches the tensile strength of the material. Thus, it is in agreement with the model by \citet{Skorov2012} that dust emission is localized.
\par
Inactive regions are (according to the model by \citet{Skorov2012}, which we have confirmed in this paper) marked by either a too thick layer of dust-aggregates so that the amount of energy transported through the dust layer is too small to cause the required outgassing rate, or by a tensile strength of the dust layer that exceeds the sublimation pressure of water ice (parts of the curves right of the maxima in Fig. \ref{cometarysurface}). While the further cause of inactivity can only dominate when the comet's distance to the Sun increases with time, the latter can be happening if the dust aggregates are too small or if the dust layer is consolidated. The latter can, for instance, happen by impacts that are energetic enough to destroy the dust aggregates. During the 4.5 billion years of the existence of the icy planetesimals before they become comets, i.e. in the Kuiper belt or in the Oort cloud, these bodies may have undergone mutual collisions and have been exposed to impacts from smaller bodies. \citet{Beitzetal2013} have experimentally shown that for impact velocities $\lesssim 1 \mathrm{km \, s^{-1}}$, the impact consolidation pressure can well be described by $p_{\mathrm{imp}} \sim \rho \, v_{\mathrm{imp}}^2$, with $v_{\mathrm{imp}}$ being the impact velocity. From the impact-consolidation work of \citet{Guettleretal2009} and \citet{Beitzetal2013}, we can thus conclude that low-velocity impacts into bodies consisting of pure dust aggregates with velocities on the order of $v_{\mathrm{imp}} \sim 10 \, \mathrm{m \,s^{-1}}$ lead to an increase in tensile strength of $p_{\mathrm{tensile}} \sim \mathrm{kPa}$, whereas higher relative collision speeds ($100 \, - \, 1000 \, \mathrm{m \, s^{-1}}$) result in materials compacted to tensile strengths of $p_{\mathrm{tensile}} \, \gg \, \mathrm{MPa}$ \citep{Beitzetal2013}. Typically, these values are also inferred for the disintegration of large meteoroids producing meteorites \citep{Petrovic2001,Popova2011}. Due to the higher surface energy of water ice, ice-dust mixtures require slightly higher impact speeds to lose their aggregate nature. It is interesting to note that the volume that is affected by these impacts is approximately the volume of the impactor \citep[see][and references therein]{Beitzetal2013} so that at a certain depth the original state of the cometary matter is retained. Thus, compacted patches have relatively strong internal cohesion so that they can only be released intact once the surroundings have been eroded to a depth exceeding that of the compressed region. This could possibly explain the emission of consolidated chunks of material exceeding the size of the dust aggregates (see Sec. \ref{sec:comcomp}). This hypothesis is supported by observations of cometary trails \citep[see, e.g.,][]{Reachetal2000,Agarwaletal2010} and of the inner coma of comet 103P/Hartley performed during the EPOXI mission\citep{AHearnetal2011}, which have shown that relatively big aggregates ($> 1 \, \mathrm{cm}$) are present in the vicinity of cometary nuclei.
\par
In general, the ejection of stronger material is possible by the increase of pressure beneath the dust layer. \citet{Hartmann1993} has experimentally shown that pressures well above $\sim 2-20 \, \mathrm{mbar}$ are required beneath the dust layer to explain the typical expansion velocities of cometary jets ($100 \, \mathrm{m \, s^{-1}}$ to $1000 \, \mathrm{m \, s^{-1}}$). However, the required high pressures can only be reached by the evaporation of water ice at temperatures exceeding $260-289 \, \mathrm{K}$, or by other, more volatile materials.

\section{\label{sec:conclusion}Conclusion}
A wealth of laboratory and microgravity experiments on dust-aggregate collisions have led to a collision model for protoplanetary dust \citep{Guettler2010}. With this dust-aggregate collision model, the growth of millimeter to centimeter-sized dust aggregates in protoplanetary disks can be explained. Ice should behave qualitatively similar, but quantitative differences are expected because (i) ice is stickier than silicates \citep{Gundlach2011b} and (ii) the collision velocities are different at 30 AU \citep{Weidenschilling1997}. As we do not have an analog of the dust-aggregate collision model for ice aggregates, it is unclear whether the expected sizes of the ice aggregates exceed those of the dust aggregates or not. For our model, this is, however, of minor importance, because we assume that dust and ice aggregates are separated inside the comet and the ice sublimation leaves the dust aggregates intact. This is consistent with meteoroid survival in interplanetary space for thousands of years \citep{Jenniskens1998,Sykesetal2004,Trigoetal2005}. However, in our model the ice-aggregate sizes cannot exceed the dust-aggregate sizes by much, because after ice sublimation, a dust layer of thickness of the ice-aggregates forms, which, once closed, might be too thick to sustain cometary activity if the ice aggregates were too big (see Sect. \ref{sec:application}).
\par
As was shown in Sect. \ref{sec:introduction}, models of the formation of planetesimals/cometesimals require either collective particle effects (e.g., the streaming instability and gravitational instability), a continuous fragmentation-agglomeration cycle, or more sticky materials than silicates (e.g., ices or aggregates-of-aggregates). If comet nuclei are leftover cometesimals, whose formation is described by one of the two competing planetesimal-formation models (gravitational instability or mass transfer), then only very gentle processes, i.e. the gravitational-instability model, can explain the gas and dust activity (i.e. the very small required tensile strengths) of comets. The mass-transfer model predicts comets with much too high tensile strengths. For the gravitational-instability formation scenario of cometesimals, the model by \citep{Skorov2012} predicts the correct tensile strength as proven by our experiments (see Sect. \ref{sec:lab}). Our experiments confirm that if the comet nuclei consist of dust (and ice) aggregates, the cohesion-induced tensile strength of the ice-free surface layer is $p_{\rm tensile} \sim 1 ~ \mathrm{Pa}~ (s/{\mathrm{1 mm}})^{-\alpha}$, with $s$ being the radius of the dust aggregates and $\alpha = 2/3$.
\par
Our experiments also show that hydrostatic compression of layers of dust aggregates leads to an additional consolidation and, thus, an increase of tensile strength towards the comet-nucleus center (see Sect. \ref{sec:lab}). Impact-induced consolidation \citep{Beitzetal2013} can also lead to an increased tensile strength. Both processes might explain the extinction of old comets \citep{Jewitt2008}, if there is no re-organization of dust aggregates after the ice has evaporated.

\section{\label{sec:future}Future Work}
The EPOXI mission to comet 103P/Hartley discovered the release of water-ice particles with sizes ranging from $1 \, \mathrm{cm}$ to $15 \, \mathrm{cm}$ at a heliocentric distance of $1 \mathrm{AU}$ \citep{AHearnetal2011}. These observations are in qualitative agreement with our results (see crosses in Fig. \ref{min_size}). The higher specific surface energy of water ice (compared to $\mathrm{SiO_2}$) implies a higher tensile strength of water ice aggregates and, therewith, a higher static pressure needed to destroy a layer composed of water-ice aggregates. Thus, changing the material from $\mathrm{SiO_2}$ to water ice increases the critical radius of the aggregates (see squares and diamonds in Fig. \ref{min_size}). It is obvious that future experiments will have to concentrate on icy agglomerates and determine their tensile strengths.
\par
Since the KOSI experiments \citep[see, e.g.,][for details]{Gruen1991}, performed about 25 years ago, no experimental study has dealt with the physics of the release of dust particles from evaporating ices. Thus, the conditions and detailed physical processes of dust emission are still not fully understood. It is mandatory to investigate the behavior of ice-dust mixtures under energy input to study in more detail the exposure and detachment of dust particles from retreating ice surfaces. Another assumption in the model by \citet{Skorov2012} is the formation of an ice-free layer of dust aggregates, which was also the basis of our experimental work presented here. It is empirically unproven whether or not a mixture of icy and dusty agglomerates leaves a layer of intact dust aggregates behind when the ice has evaporated. This is also a point to be addressed in future experimental investigations.
\par
Certainly some other physical processes also need to be better studied. For example, the key effect of cosmic irradiation, thermal gradients and collisions in comets were pointed out by \citet{KruegerKissel2006} among others. The carbon-rich, fine-grained and fluffy matrix forming comets might be affected by these processes. When losing much of their oxygen and hydrogen due to released water vapor, the remaining solid phase of the organic material tends to form polymers or clathrates \citep{KruegerKissel2006}. The stickiness of these materials could influence the dust release and should be studied in future experiments. Besides laboratory experiments, we hope that significant insight in this regard will be achieved by the different experiments planned on the Philae lander of the Rosetta mission \citep[see, e.g.,][]{Goesmannetal2009, Spohnetal2009}. Rosetta will be a key instrument to shed light on the formation processes of comets, because the underlying assumption of a gravitationally-bound clump of dust and ice aggregates \citep{Skorov2012} can be directly tested on comet 67P/Churyumov-Gerasimenko.  Particularly, the SESAME experimental package on-board the Philae lander \citep{Seidenstickeretal2007} can help to better understand the structure and composition of the cometary surface material by measuring soundwave propagation in the material (CASSE), or the response of the material to induced high-frequency currents (PP). If successful, these experiments are capable to measure physical properties of the surface material, like the Young's modulus, the Poisson number, and the specific resistivity.
\par
Our experiments showed that static compression alters the contact forces between the dust aggregates. Although our Figure \ref{fig3} indicates that the two-minute compression applied to the aggregates are sufficient to result in a saturation of the tensile strength, there could be some flowability of the aggregate ensemble on (much) longer timescales, which might turn our to be important for the modeling of comet nuclei. This also needs to be investigated in the future. As above, ice aggregates are basically unknown in their mechanical behavior so that these measurements need also be performed for pure ice and mixtures of ice and dust agglomerates.
\par
In this work, we have concentrated on the gravitational-instability model of cometesimals. An alternative cometesimal-formation model was presented by \citet{Kataokaetal2013}. Based on the enhanced stickiness of sub-micrometer-sized ice particles, the formation of icy cometesimals by direct sticking seems possible. It is unclear what implications in terms of tensile strength of the forming body this model has, particularly, if an admixture of dust particles is taken into account. This will have to be addressed in future studies as well.

\subsection*{Acknowledgements}
We thank the two anonymous reviewers for their very detailed and extremely helpful comments, which helped improving this manuscript considerably, Yuri Skorov and Horst Uwe Keller for valuable discussions, and Joshua Steyer for helping us in carrying out the laboratory experiments. The experiments were funded through DLR grant 50WM1236.

\bibliographystyle{model2-names}
\bibliography{bib}

\end{document}


\begin{frontmatter}
\title{Corrigendum to "Comets formed in solar-nebula instabilities! -- An experimental and modeling attempt to relate the activity of comets to their formation process" (Icarus 235, 156-169, 2014)}


\author[igep]{J. Blum}
\author[igep]{B. Gundlach}
\author[igep]{S. Mühle}
\author[ieec]{J.M. Trigo-Rodriguez}
\address[igep]{Institut für Geophysik und extraterrestrische Physik, Technische Universität Braunschweig, \\Mendelssohnstr. 3, D-38106 Braunschweig, Germany}
\address[ieec]{Institute of Space Sciences (CSIC), Campus UAB, Facultat de Ciències, \\Torre C-5 pares, 2a pl., 08193 Bellaterra (Barcelona), Spain}
\end{frontmatter}

\setlength{\tabcolsep}{10pt}
\renewcommand{\arraystretch}{2}
\renewcommand{\topfraction}{1.0}
\renewcommand{\bottomfraction}{1.0}
In our recent paper \citep{Blum2014}, we studied how different formation scenarios determine the physical properties of cometary nuclei. One of our main conclusions implied that comets must have formed by gravitational instability in an ensemble of dust aggregates in order to explain their water-driven dust activity at heliocentric distances around $\sim1\,\mathrm{AU}$. The ejection of dust aggregates from the cometary surface is only possible if the resulting gas pressure beneath the aggregate layer exceeds the tensile strength of the material, which can be very low ($\sim 1 \, \mathrm{Pa}$, in the case of formation by gravitational instability). We used a thermophysical model to calculate the resulting gas pressure beneath aggregate layers with variable thickness at different heliocentric distances \citep[Fig. 9 in][]{Blum2014} in order to test whether water-driven dust activity is possible. The pressure at the dust-ice interface was determined by computing the temperature of the dust-covered ice surface by numerically solving the heat transfer equation and by taking the permeability of the dust-aggregate layer into account.
\par
Unfortunately, we made a simple mistake in the numerical treatment of the thermophysical model by misplacing the decimal point of Boltzmann's constant. This error only affects Figs. 9 and 10 in \citet{Blum2014} where we derived the minimum dust-aggregate size that can be ejected by water-ice activity. We corrected the numerical error in our thermophysical model and recalculated the resulting pressures beneath the aggregate layers and the minimum dust-aggregate sizes that can be released due to the outgassing of water ice.
\par
\begin{figure}[b!]
\centering
\includegraphics[angle=90,width=1\columnwidth]{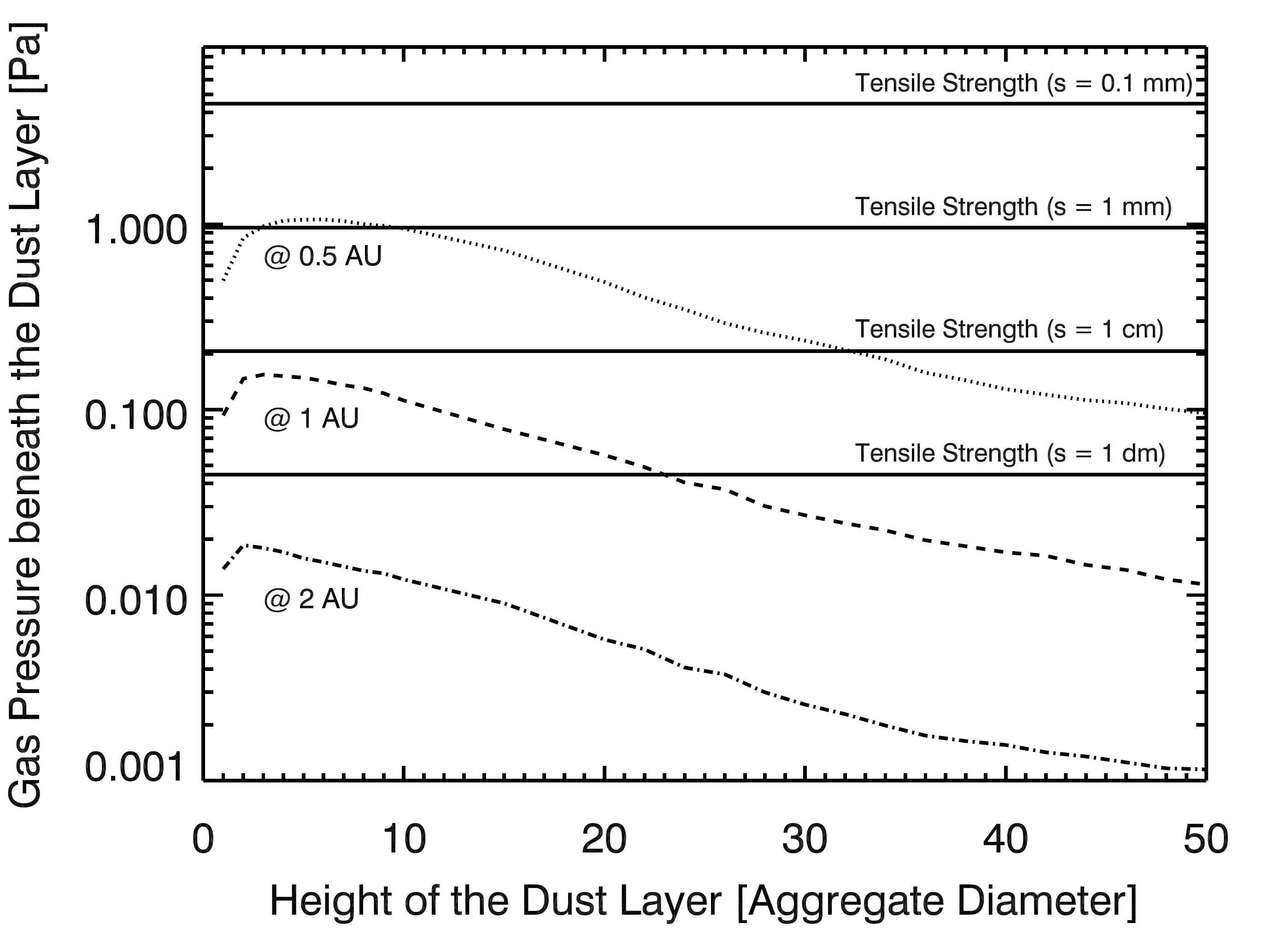}
\caption{Updated version of Fig. 9 in \citet{Blum2014} showing the recalculated gas pressure at the dust-ice interface as a function of the thickness of the covering dust-aggregate layers, in units of the dust-aggregate diameter. The three curves were calculated for dust-aggregate radii of $s=1\, \mathrm{mm}$, but are virtually independent of this size. The calculations were performed for three different heliocentric distances of $0.5 \,\mathrm{AU}$ (dotted curve), $1 \,\mathrm{AU}$ (dashed curve), and $2 \,\mathrm{AU}$ (dashed-dotted curve), respectively. For comparison, the tensile strengths of dust-aggregate layers \citep[solid lines; see Eq. 1 in][]{Blum2014} are shown for different dust-aggregate radii $s=0.1\, \mathrm{mm}$, $s=1\, \mathrm{mm}$, and $s=1\, \mathrm{cm}$ and $s=1\, \mathrm{dm}$), respectively.}
\label{fig1}
\end{figure}

Fig. \ref{fig1} \citep[updated version of Fig. 9 in][]{Blum2014} shows the recalculated gas pressures at the dust-ice interface derived for a dust-aggregate radius of $1 \, \mathrm{mm}$ and for three different heliocentric distances (dotted, dashed and dashed-dotted curves for $0.5 \,\mathrm{AU}$, $1 \,\mathrm{AU}$ and $2 \,\mathrm{AU}$, respectively). The calculations have shown that the resulting gas pressures are virtually identical for different dust-aggregate sizes. Thus, we decided to only show the  $1 \, \mathrm{mm}$ cases here. The recalculated gas-pressure curves are lower by almost one order of magnitude than the old curves. A decrease of the gas pressure at the dust-ice interface implies an increase of the minimum radius of those dust aggregates that can be released by the gas pressure, because the tensile strength of dust-aggregate layers (solid horizontal lines in Fig. \ref{fig1}) decreases with increasing dust-aggregate size \citep[see Eq. 1 in][]{Blum2014}. For example, the minimum dust-aggregate radius increased from $0.6 \, \mathrm{mm}$ to $15.5 \, \mathrm{mm}$ at $1 \, \mathrm{AU}$ compared to the old calculations.
\par
\begin{figure}[b!]
\centering
\includegraphics[angle=90,width=1\columnwidth]{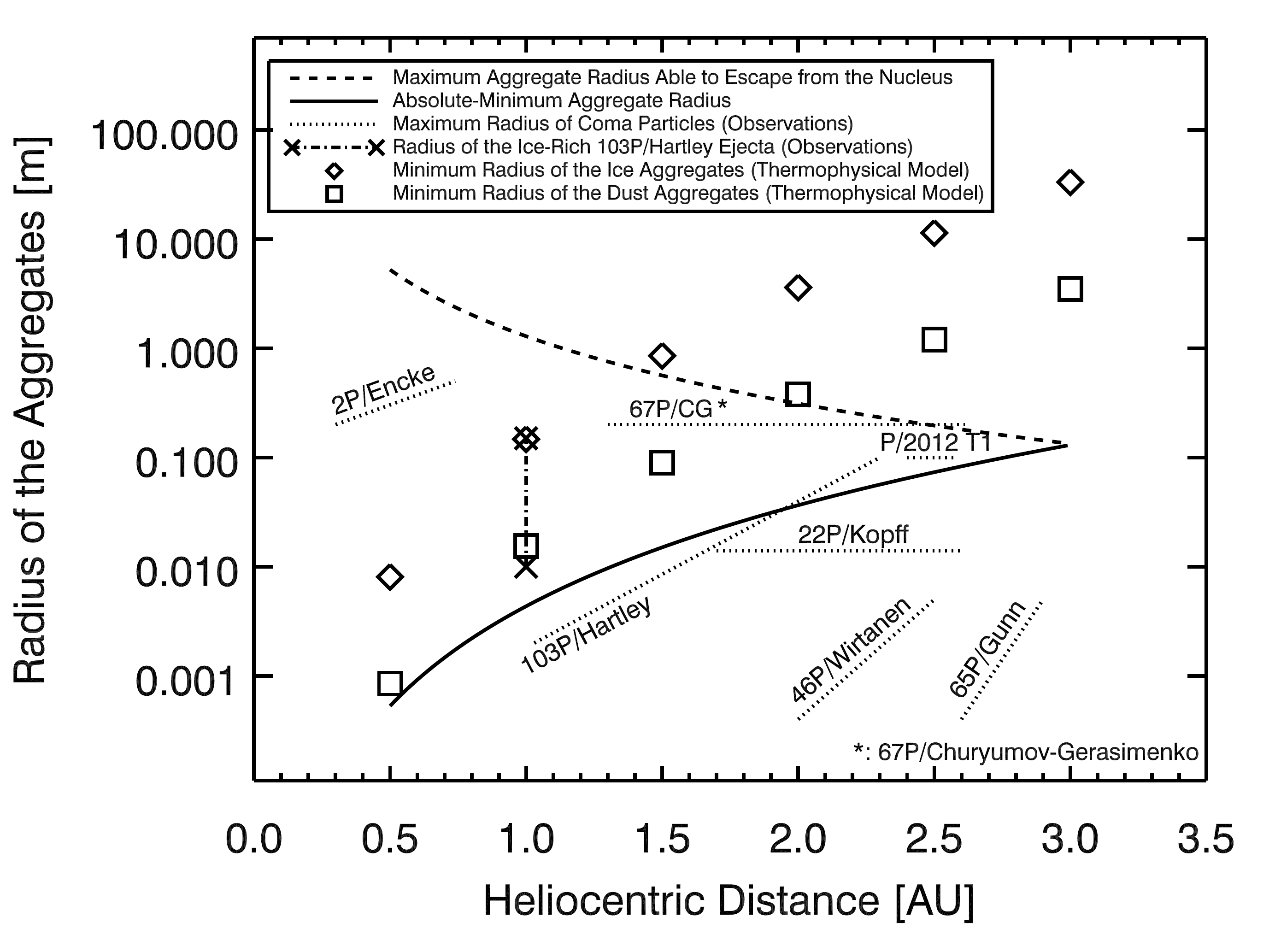}
\caption{Updated version of Fig. 10 in \citet{Blum2014} presenting the minimum radius of dust aggregates (squares) and ice aggregates (diamonds) that can be released by water-ice sublimation as a function of heliocentric distance. Dust layers composed of aggregates larger than these sizes are released, because the static gas pressure exceeds the tensile strength of the dust-aggregate layer. The solid curve shows the absolute-minimum dust-aggregate radius that can be released, derived by assuming that the absorbed solar energy is completely used for the sublimation process. Additionally, the maximum dust size that can leave a 1-$\mathrm{km}$-sized comet nucleus due to gas friction is shown by the dashed curve (for details, see text). For comparison, the maximum grain radii of coma particles derived from observational data of the following comets are shown by the dotted lines: 2P/Encke \citep{Epifani2001}, 103P/Hartley \citep{Epifani2001}, 67P/Churyumov-Gerasimenko \citep{Moreno2004}, 22P/Kopff \citep{Moreno2012}, 46P/Wirtanen \citep{Colangeli1998}, 65P/Gunn \citep{Colangeli1998}, and P/2012 T1 \citep[Panstarrs;][]{Moreno2013}. Additionally, the estimated sizes of ejected ice-rich particles from the nucleus of comet 103P/Hartley during the flyby of the EPOXI spacecraft \citep[][]{AHearnetal2011} are visualized by the crosses and the dashed-dotted line.}
\label{fig2}
\end{figure}
In order to take these corrections fully into account, we also reworked the old Fig. 10 in \citet{Blum2014}, which shows the minimum radius of the dust aggregates that can be released by water-ice sublimation as a function of heliocentric distance. The new version of this figure (Fig. \ref{fig2}) presents the recalculated minimum radii of the dust aggregates for different heliocentric distances (squares). For a comparison, we additionally derived the absolute-minimum dust-aggregate radius that can be released (solid curve) by assuming that the absorbed solar power is completely converted into sublimation of water ice at the ice-dust interface (i.e., no thermal emission from the dust surface). The deviation of the black curve from the squares directly shows the importance of thermal emission on the energy budget of the system. For small heliocentric distances, the energy consumed by thermal emission is relatively low, i.e., most of the energy received is transported through the dust-aggregate layer to the water-ice surface where the energy is used for sublimation of the water ice. For high temperatures of the aggregate layer (i.e., at short heliocentric distances), the heat conductivity through the dust-aggregate layer is high due to radiative transport inside the pores of the dust layer \citep[$\lambda \propto T^3$, where $\lambda$ is the heat conductivity and $T$ is the temperature;][]{Gundlach2012}. This enables an effective heat transport through the dust-aggregate layer. At larger heliocentric distances, however, the radiative heat conductivity decreases so that the ratio of energy available for the sublimation process to energy available for thermal emission decreases with increasing heliocentric distance (i.e., the black curve deviates more and more from the squares with increasing heliocentric distance).
\par
The dashed curve in Fig. \ref{fig2} shows the maximum dust size that can leave a 1-$\mathrm{km}$-sized comet nucleus due to gas friction. For its derivation, we assumed that the outward-directed gas-friction force $F_\mathrm{out} = \pi \, s^2 \, p_\mathrm{max} \, (R/r)^2$ has to exceed the inward-directed gravitational force $F_\mathrm{in} = m \, G \, (M/R^2) \, (R/r)^2$, i.e. $F_\mathrm{out} \ge F_\mathrm{in}$. Here, $s$, $p_\mathrm{max}$, $R$, $r$, $m$, $G$, and $M$ are the radius of the released dust aggregate, the maximum gas pressure achievable at the surface of the comet nucleus, the radius of the comet nucleus, the distance to the center of the comet nucleus, the mass of the dust aggregate, the gravitational constant, and the mass of the comet, respectively. The $1/r^2$ dependence of the drag force is due to the geometrical decrease of outflowing water molecules per unit area. The maximum gas pressure of the evaporating water ice can be calculated by assuming that the absorbed Solar energy is completely used for the sublimation process. Inserting the mass density of the dust aggregates, $\rho_\mathrm{agg}$, and the comet nucleus, $\rho$, and assuming spherical dust aggregates and comet nuclei, one yields the emission condition for the size of the dust aggregates,
\begin{equation}\label{eq:maxsize}
    s \le \frac{3}{16 \pi} \frac{p_\mathrm{max}}{G \rho_\mathrm{agg} \rho} \frac{1}{R} \, ,
\end{equation}
which is independent of the distance to comet nucleus. Please mind that $p_\mathrm{max}$ is a function of heliocentric distance $a$. For the maximum ejectable particle size shown in Fig. \ref{fig2}, we used $p_\mathrm{max} = 0.35 \mathrm{Pa} \, (\mathrm{1 AU}/a)^2$, $R=1$ km, $\rho_\mathrm{agg} = 1500 \mathrm{kg\,m^{-3}}$, $\rho = 500 \mathrm{kg\,m^{-3}}$. At short heliocentric distances, the gas friction is sufficient to drag away $\mathrm{m}$-sized dust particles. This maximum size decreases further away from the Sun due to a decrease of the evaporation rate. The region between the solid and the dashed curves, thus, provides the size range of dust aggregates present in the cometary coma. This region is for a 1-km-sized nucleus relatively narrow at heliocentric distances between $2\, \mathrm{AU}$ and $3\, \mathrm{AU}$ but considerably broadens closer to the Sun. For larger comet nuclei, the maximum dust size decreases inversely proportional to the size of the nucleus according to Eq. \ref{eq:maxsize} so that a comet with $R=10$ km cannot emit dust beyond $\sim 2$ AU due to the outgassing of water ice alone.
\par
In addition, we also calculated the critical radius of water ice aggregates that can be released due to water-ice activity (diamonds) by assuming a tenfold higher tensile strength due to the higher specific surface energy of water-ice compared to silica dust \citep[see][for details]{Gundlach2011b}. For comparison, data of the dust-particle sizes observed in cometary comae (dotted lines) and the estimated sizes of the ejected ice-rich particles from the nucleus of comet 103P/Hartley during the flyby of the EPOXI spacecraft \citep[crosses; ][]{AHearnetal2011} are shown. From Fig. \ref{fig2}, it becomes obvious that cometary dust activity beyond $\sim 2 \,\mathrm{AU}$ cannot be provided by the evaporation of water-ice alone but needs to be augmented by the evaporation of ices more volatile than water ice.

\subsection*{Acknowledgements}
We thank Yuri Skorov for bringing to our attention the energy-budget problem in the thermophysical model of our original paper.

\bibliographystyle{model2-names}
\bibliography{bib}